\documentclass[fleqn,usenatbib]{mnras}
\usepackage[switch]{lineno} 
\usepackage{newtxtext,newtxmath}
\usepackage{amsmath, bm} 
\usepackage{hyperref}
\usepackage{color}
\usepackage{subfigure}
\usepackage[T1]{fontenc}

\DeclareRobustCommand{\VAN}[3]{#2}
\let\VANthebibliography\thebibliography
\def\thebibliography{\DeclareRobustCommand{\VAN}[3]{##3}\VANthebibliography}

\usepackage{graphicx}	% Including figure files
\usepackage{amsmath}	% Advanced maths commands

\title[Fast outflow in MAD]{Fast Outflow Driven from the Magnetically Arrested Disc}

\author[Yi and Li]{
Taili Yi$^{1}$,
and Jia-Wen Li$^{1}$,\thanks{E-mail: jwliynu@ynu.edu.cn}
\\
$^{1}$Department of Astronomy, School of Physics and Astronomy, Key Laboratory of Astroparticle Physics of Yunnan Province, Yunnan University,\\ Kunming 650091, China
}

\date{Accepted 2026 October 17. Received 2026 September 30; in original form 2026 June 17}

\pubyear{\the\year{}}

\begin{document}
\label{firstpage}
\pagerange{\pageref{firstpage}--\pageref{lastpage}}
\maketitle

\begin{abstract}
Fast outflows are commonly observed in accreting black hole systems, yet their launching mechanism remains debated, especially when the gas is highly ionized and radiation line driving becomes inefficient. We investigate magnetically driven outflows from magnetically arrested discs (MADs) using a steady, axisymmetric magnetocentrifugal outflow model. The outflow equations follow the classical trans-magnetosonic MHD framework, while the disc structural scalings and outflow-base conditions are motivated by general relativistic magnetohydrodynamics (GRMHD) simulations. We solve for the trans-magnetosonic outflow solutions along prescribed large-scale poloidal magnetic field lines. The outflows launched from the inner disc are strongly accelerated and the formal Newtonian solutions can enter the mildly relativistic regime. The outflow velocity decreases systematically with increasing footpoint radius, producing a radially stratified outflow in which inner streamlines reach UFO-like velocities while outer streamlines form slower components. We also find that the MAD solutions have a large magnetic lever arm $\tilde{l}=(R_{\rm A}/R)^2$ is typically several tens and can reach $\sim10^2$, which implies that the outflow can carry a large amount of the specific angular momentum of the disc material at its footpoint, and we find that about $55\%$ of the angular-momentum estimated between the outer and inner disc boundaries is removed by the magnetic outflow. The total mass-loss rate in the outflow is about $14\%$ of the disc mass-accretion rate in our fiducial model. These results suggest that the MADs provide a natural environment for producing both fast, radially stratified disc outflow and efficient disc angular-momentum removal. MAD-driven outflow may therefore provide a magnetic origin for ultra-fast absorbers in low-luminosity active galactic nuclei (AGNs) and X-ray binaries (XRBs), while also contributing significantly to the angular-momentum transfer of the accretion flow.
\end{abstract}



%%%%%%%%%%%%%%%%%%%%%%%%%%%%%%%%%%%%%%%%%%%%%%%%%%

%%%%%%%%%%%%%%%%% BODY OF PAPER %%%%%%%%%%%%%%%%%%

\section{Introduction}\label{sec:introduction}
Black hole accretion flows release gravitational energy not only through radiation, but also through collimated jets and less-collimated disc winds/outflows. Such outflows are observed across the black hole mass scale, from X-ray binaries to active galactic nuclei. These fast outflows carry a large amount of angular momentum and energy, which play an important role in the study of black hole binary state transition, coevolution of a supermassive black hole and its host galaxy, and AGNs feedback \citep[][]{2014A&A...562A..21C,2014ARA&A..52..529Y,2015Natur.519..436T,2015ARA&A..53..115K,2021NatAs...5...13L}. Over the past several years, high-ionization blue-shifted absorption lines have been observed in the X-ray spectra of approximately $40\%$ of AGNs in our local universe, the velocity of the outflows ranges from $0.03c$ to $0.5c$ with ionization parameter $\xi \sim10^{2}-10^{6}\ \rm{erg \ cm \ s^{-1}}$. These fast outflows are believed to be generated from the accretion disc very close to the central black hole \citep[][]{2010A&A...521A..57T,2015ApJ...805...17F,2018ApJ...857...34Z}. These fast outflows can significantly affect the dynamics, density, temperature profiles of the accretion disc, and subsequently change the observed spectrum \citep[][]{1994ApJ...433..746S,2013ApJ...765..149C,2013ApJ...767...30B,2014ApJ...786....6L,2018ApJ...857...34Z,2019ApJ...872..149L,2023Sci...381..961Y}, thus the winds/outflows launching and acceleration from AGNs, black hole X-ray binaries are of great interest. Among these fast outflows, a new class of the outflows, named ultra-fast outflows (UFOs) have drawn much attention due to its distinctive physical properties. These outflows are accelerated to near-relativistic speed around $0.1c-0.7c$, with a nearly Compton-thick H-equivalent column density of $N_{\rm H} \gtrsim 10^{23} \rm{cm ^{-2}}$, the ionization state of the gas is usually very high with the ionization parameter $\log \xi \gtrsim 4$ (where $\xi=L_{\rm ion}/{nr^2}$ , $L_{\rm ion}$ is the ionizing luminosity, $n$ the plasma density and $r$ the radial distance from the ionizing source). UFOs appear to be present not only in AGNs \citep[][]{2009ApJ...701..493R,2010A&A...521A..57T,2010ApJ...719..700T,2011MNRAS.418L..89T,2013MNRAS.430.1102T,2015Natur.519..436T,2015MNRAS.451.4169G,2017MNRAS.469.1553P,2018A&A...610L..13D,2020ApJ...895...37R}, but also in some XRBs \citep[][]{2006Natur.441..953M,2008ApJ...680.1359M,2012ApJ...759L...6M,2014ApJ...784L...2K,2018MNRAS.479.3978K,2021ApJ...906...11W} and presumably in the ultraluminous X-ray sources \citep[][]{2016ApJ...826L..26W,2018MNRAS.479.3978K}

The acceleration mechanisms of these fast outflows, however, are still under debate. There are two main models which are believed to be responsible for the launching of these outflows: the magnetic centrifugally driven and the radiation driven (mainly the radiation UV line-force driven). The line-force driven mechanism has been extensively investigated in the literature through both analytical modeling and numerical simulations \citep[][]{1995ApJ...454L.105M,2000ApJ...543..686P,2014MNRAS.438.3024L,2021MNRAS.503.1442M,2021ApJ...922..262Y,2026A&A...705A...4B}. This mechanism is efficient in driving fast outflows with low ionization state in the luminous black hole accretion systems. If one taking the high ionization state (most UFOs are so highly ionized that there is little UV or soft X-ray opacity), special relativity effect, and radiation-drag effect into consideration, the radiation force becomes inefficient in driving fast outflows with high inclination angle in some low-luminosity systems \citep[][]{,2014MNRAS.443.2154T,2014ApJ...789...19H,2015MNRAS.446..663H,2020A&A...633A..55L,2021A&A...646A.111L,2021MNRAS.501.4850R,2022MNRAS.515.5594W,2023A&A...670A.122M}. Alternatively, these fast outflows can be centrifugally accelerated by the large-scale magnetic field anchored in the accretion disc \citep[][]{1982MNRAS.199..883B,2010LNP...794..233S,2010ApJ...715..636F,2010A&A...512A..82M,2013ApJ...767...30B,2014ApJ...786....6L,2014ApJ...793...31S,2018ApJ...857...34Z,2019ApJ...872..149L}, which provide a plausible means of efficient acceleration as observed \citep[][]{1994ApJ...429..139C,1994ApJ...434..446K,2006Natur.441..953M,2015ApJ...814...87M,2015ApJ...805...17F,2017NatAs...1E..62F,2018ApJ...853...40F,2018ApJ...852...35K,2020ApJ...904...30M,2021ApJ...906...11W,2023A&A...670A.182M}. The large-scale magnetic field threading the accretion disc may be formed through the advection of the weak external magnetic field by the accreting gas and/or an in situ generation by the turbulent dynamo \citep[][]{1981MNRAS.195..897P,1998ApJ...500..703R,2003ApJ...592.1042I,2013ApJ...765..149C,2013ApJ...767...30B,2014ApJ...796...29S,2019ApJ...872..149L,2020MNRAS.494.3656L,2020MNRAS.492.1855M}. 

The efficiency of magnetic launching outflow depends significantly on the strength and geometry of the large-scale magnetic flux threading the accretion disc. A particularly important regime is the magnetically arrested disc (MAD), in which the accumulated poloidal magnetic flux becomes dynamically important and partially arrested the accretion flow. The study of MAD has transitioned from a theoretical curiosity into a cornerstone of contemporary  accretion astrophysics \citep[][]{1974Ap&SS..28...45B,2003PASJ...55L..69N,2011MNRAS.418L..79T,2022ApJ...924..124Y,2023Sci...381..961Y}. MADs have become an important paradigm in black hole accretion theory because they can produce powerful relativistic jets through the Blandford–Znajek mechanism, with jet efficiencies that may exceed the accretion power \citep[][]{1977MNRAS.179..433B,2011MNRAS.418L..79T,2012MNRAS.423.3083M}. Observationally, dynamically important magnetic fields near black holes are supported by polarimetric Event Horizon Telescope observations of M87* and Sgr A*, as well as by multiwavelength evidence for a MAD-like state in the black hole X-ray binary MAXI J1820+070  \citep[e.g.,][]{2013Natur.501..391E,2014Natur.510..126Z,2019ApJ...875L...5E,2021ApJ...910L..13E,2022ApJ...930L..16E,2023Sci...381..961Y}, dynamically important large-scale magnetic fields exist around both supermassive black holes and stellar-mass black holes.

Although MADs are often discussed in the context of relativistic Blandford-Znajek jet \citep[][]{1977MNRAS.179..433B}, they should also provide an ideal environment for launching powerful disc winds/outflows. In a MAD, the disc surface is threaded by strong, ordered poloidal magnetic fields, and the magnetic pressure can become comparable to or larger than the gas pressure. These conditions naturally favor magnetocentrifugal acceleration and may produce wide-angle outflows in addition to collimated jets. Recent numerical simulations show that MADs can drive fast outflow and the related magnetic flux eruptions may provide an efficient channel for vertical angular momentum transport \citep[][]{2022ApJ...941...30C}. In thick MADs, disc winds/outflows may even remove angular momentum at a rate comparable to turbulent stresses within the disc \citep[][]{2024ApJ...965..175M}. MAD-driven winds/outflows may play an important role in regulating accretion and generation of powerful, wide-angle, sub-relativistic outflows. Unlike Standard and Normal Evolution (SANE) discs, where magnetic fields usually weak and maybe turbulent, the coherent magnetic field in a MAD system is strong enough to dominate the gas dynamics at the disc-outflow interface. The gas around the disc surface is flung outward along the stiff, rotating poloidal field lines, and then magnetic centrifugally accelerated to form the fast/ultra-fast outflows under the Blandford-Payne mechanisms \citep[][]{1982MNRAS.199..883B}. 

In this work, we construct a semi-analytic model for magnetically driven outflows launched from the surface of a MAD. Motivated by the radial structure found in general relativistic magnetohydrodynamics (GRMHD) simulations done by \cite{2022ApJ...941...30C}, we prescribe the disc density, angular velocity, magnetization, and large-scale poloidal magnetic field geometry, and then solve for steady axisymmetric outflows along magnetic field lines. We focus on the trans-magnetosonic solutions and examine how the outflow accelerates through the slow magnetosonic point, the Alfv\'en point, and the fast magnetosonic point. We show that a MAD-like magnetic configuration can accelerate disc-surface gas to sub-relativistic or mildly relativistic velocities, producing a radially stratified outflow that may be relevant to UFOs in luminosity AGNs and XRBs. In Section \ref{sec:model}, we present the semi-analytic model for magnetically driven outflows and derive the Bernoulli equation along a field line. Section \ref{sec:NuSet} contains the adopted MAD disc structure and related numerical setup. The results of the numerical solutions and discussions are presented in Sections \ref{sec:results} and \ref{sec:CandD}. 

\section{Model}\label{sec:model}
We consider steady axisymmetric outflows launched from the surface of a MAD disc, (the base of the outflow), and we shall work in cylindrical coordinates $(R,\varphi,z)$, i.e., $\partial/\partial t = \partial/\partial \varphi = 0$. As we know that the dynamical structure of the accretion flow, the large-scale field configuration of the disc, and the properties of the outflows are a tightly coupled system. To derive the dynamical structure of the disc, for instance, one needs to know the outflow properties first, while the calculations of the outflow solution require the specific field configuration/strength above the disc, which in turn is decided by the whole structure of the disc itself \citep[see][for details]{2019ApJ...872..149L}. To focus on the dynamics of the fast outflows, for simplicity, we adopt a self-similar field configuration as that suggested in \cite{1994A&A...287...80C} this is also valid in some of the numerical simulations \citep[][]{2015ApJ...804..101Y,2018ApJ...857...34Z,2020MNRAS.492.1855M}. We adopt the magnetic field configuration as
\begin{equation}\label{eq:Bz_midplane}
    B_{\rm z}(R,z=0)=\left[1+\left(\frac{R}{R_{\rm in}}\right)^2\right]^{-1/2},
\end{equation}
where $R_{\rm in}$ is the inner radius of the accretion disc $(R_{\rm in}=6R_{\rm g}=6GM/c^2)$. Based on above boundary conditions, the stream function of an irrotational and solenoidal magnetic field above the disc can be written as 
\begin{equation}\label{eq:magnetic flux} 
    \Phi = \left[ \left( \frac{R}{R_{\rm in}} \right)^2 + \left( 1 + \left| \frac{z}{R_{\rm in}} \right| \right)^2 \right]^{1/2} - \left( 1 + \left| \frac{z}{R_{\rm in}} \right| \right).
\end{equation}
The strength and configuration of the large-scale poloidal magnetic field are then calculated by
\begin{equation}\label{eq:field components}
\begin{aligned}
B_{\rm R}(R,z) &= -\frac{1}{R} \frac{\partial \Phi}{\partial z},\\
B_{\rm z}(R,z) &= \frac{1}{R} \frac{\partial \Phi}{\partial R}.
\end{aligned}
\end{equation}

The outflow is accelerated along above magnetic field, the dynamics of which can be described by a set of magnetohydrodynamics equations which reads \citep[][]{1996ASIC..477..249S},
\begin{equation}\label{eq:MHD}
\begin{aligned}
        \nabla \times \left(\pmb{v}\times\pmb{B}\right)=&0,  \\
      -\nabla P - \rho \nabla \Psi_{\rm g} + \frac{1}{4\pi}\left( \nabla \times \pmb{B} \right)\times \pmb{B} = &  \rho \left( \pmb{v}\cdot\nabla \right)\pmb{v}, \\
    \nabla \cdot \left( \rho \pmb{v} \right)=&0,\\
    \nabla\cdot \pmb{B}=&0,
\end{aligned}
\end{equation}

For the dynamics of the magnetic outflows, under the axisymmetric assumption, the magnetic field and the velocity of the outflow can be decomposed into toroidal and poloidal components, i.e.,
\begin{equation}\label{eq:poloidal and toroidal components} 
    \mathbf{B} = \mathbf{B}_{\rm p} + B_{\varphi} \mathbf{e}_{\varphi}, \qquad \mathbf{v} = \mathbf{v}_{\rm  p} + v_{\varphi} \mathbf{e}_{\varphi}, 
\end{equation}
where the poloidal velocity and magnetic field are ${B}_{\rm p}$ and ${v}_{\rm  p}$, and the toroidal components are ${B}_{\varphi}$ and ${v}_{\varphi}$, respectively. The dynamics of the outflow are described by a set of conservation equations which can be derived from equation \eqref{eq:MHD}  \citep[see also][for details]{1994A&A...287...80C}, 
\begin{equation}\label{eq:induc_poloi}  
    v_{\varphi} - R\Omega = \frac{v_{\rm p}B_{\varphi}}{B_{\rm p}},
\end{equation}

\begin{equation}\label{eq:angular momentum conser}
    R v_{\varphi} - \frac{RB_{\varphi}B_{\rm p}}{4\pi\rho v_{\rm p}} = \Omega R_{\rm A}^2,
\end{equation}

\begin{equation}\label{eq:mass_conser}
    \kappa = \frac{\rho v_{\rm p}}{B_{\rm p}},
\end{equation}
\begin{equation}\label{eq:Bernoulli integral}
\frac{1}{2}v_{\rm p}^2 + \frac{1}{2}(v_{\varphi} - R\Omega)^2 + h + \Psi_{\rm eff} \equiv E.
\end{equation}
where $\Omega$ is the angular velocity of the field line at the footpoint $R_{z_{\rm h}}$. $\kappa$ and $E$ are the ratio of the mass flux to the magnetic flux, and the specific energy, respectively. Both of them are constant along each magnetic filed line (i.e., the streamline of the magnetic outflow). $R_{\rm A}$ is the Alfv\'en radius and $\Psi_{\rm eff}$ is the effective potential (a sum of the potential energy related to the centrifugal and gravitational forces along the field line) which is defined as
\begin{equation}\label{eq:efftive potential}
    \Psi_{\rm eff}(R,z) = -\frac{GM_{\rm BH}}{(R^2 + z^2)^{\frac{1}{2}}} - \frac{1}{2} R^2 \Omega^2,
\end{equation}
where $M_{\rm BH}$ is mass of the central black hole. To solve the Bernoulli equation \eqref{eq:Bernoulli integral}, one needs to know the enthalpy $h=\int dP/\rho$ of the outflow along the streamline. The outflowing gas, however,  should undergo complicated heating and cooling processes as it's being accelerated from the surface of the disc to a large distance, which is out of the scope of our present work. We employ, for simplicity, a polytropic equation of state,
\begin{equation}
    P=K \rho^\gamma,
\end{equation}
where $K$ is a constant along each field line, and $\gamma$ is the adiabatic index. With above equation of state, the  enthalpy $h$ of the outflow along each field line can be derived as 
\begin{equation}\label{eq:specific enthalpy}
    h=\frac{\gamma}{\gamma-1}K \rho^{\gamma-1}.
\end{equation}
Substituting equations \eqref{eq:induc_poloi}, \eqref{eq:angular momentum conser}, \eqref{eq:mass_conser}, \eqref{eq:efftive potential} and \eqref{eq:specific enthalpy} into equation \eqref{eq:Bernoulli integral} we obtain 
\begin{equation}\label{eq:BernEq}
    \frac{B_{\rm p}^2 \rho_{\rm A}}{8\pi \rho^2} + \frac{\Omega^2 \rho_{\rm A}^2 (R^2 - R_{\rm A}^2)^2}{2R^2(\rho - \rho_{\rm A})^2} + \frac{\gamma}{\gamma-1}K \rho^{\gamma-1} + \Psi_{\rm eff} = E.
\end{equation}
Here, the Bernoulli constant $E$ is written in the frame co-rotating with the footpoint of the magnetic field line, in the rest frame the Bernoulli constant $E_0= E + R_{\rm A}^2 \Omega^2$.

We define dimensionless quantities by
\begin{equation}\label{eq:dimensionless_paras}
\begin{aligned}
    \tilde{R}=\frac{R}{R_{\rm A}},   \tilde{\rho} =&\frac{\rho}{\rho_{\rm A}},\tilde{z} = \frac{z}{R_{\rm A}},\tilde{E} = E/\frac{GM_{\rm BH}}{R_{\rm A}},\\  
    \xi=\frac{\Theta_{z_{\rm h}}}{\beta_{z_{\rm h}}}\frac{\tilde{\rho}_{z_{\rm h}}}{\tilde{R}_{z_{\rm h}}}, f_{\rm \Omega}=&\frac{\Omega}{\Omega_{\rm K}}, \theta=\Theta_{z_{\rm h}}\frac{\gamma}{\gamma - 1}\frac{\tilde{\rho}_{z_{\rm h}}^{1-\gamma}}{\tilde{R}_{z_{\rm h}}},\beta_{z_{\rm  h}}=\rho_{z_{\rm h}}c_{{\rm s}, z_{\rm h}}^2/\frac{B_{{\rm p}, z_{\rm h}}^2}{8\pi},
\end{aligned} 
\end{equation}
where the subscripts $\rm A$ and $z_{\rm h}$ refer to the values of the quantities at the Alfv\'en point and the field footpoint. $\beta_{z_{\rm h}}$ and $\Theta_{z_{\rm h}}$ are the ratio of gas pressure to magnetic pressure and dimensionless temperature at the disc surface. $f_{\rm \Omega}$ is the ratio of the angular velocity of the footpoint to the Keplerian angular velocity. Then the Bernoulli integral, i.e., equation \eqref{eq:BernEq}, can be rewritten into dimensionless form as a function of $\tilde{R}$ and $\tilde{\rho}$,
\begin{equation}\label{eq:Bernoulli dimensionless}
    \frac{\xi}{\tilde{\rho}^2}\frac{B_{\rm p}^2}{B_{{\rm p}, z_{\rm h}}^2} + \frac{f_{\rm \Omega}^2}{2\tilde{R}^3_{z_{\rm h}}}\left[ \frac{\tilde{R}^2 - 1}{\tilde{R}(\tilde{\rho} - 1)}\right]^2 + \theta \tilde{\rho}^{\gamma - 1} - \frac{1}{(\tilde{R}^2 + \tilde{z}^2)^{\frac{1}{2}}} - \frac{f_{\rm \Omega}^2}{2}\frac{\tilde{R}^2}{\tilde{R}^3_{z_{\rm h}}}\\ = \tilde{H}(\tilde{R},\tilde{\rho})=\tilde{E}, 
\end{equation}
around the surface of the MAD disc, the magnetic field is very strong, i.e., $\beta_{z_{\rm h}}\lesssim 1.0$, and the magnetic pressure dominates over the gas pressure. The plasma is well confined on the magnetic field line and centrifugally accelerated along the field line. The solution of the function $\tilde \rho(\tilde R)$ is a contour line of the Bernoulli function $\tilde H(\tilde{R},\tilde{\rho})$ in the ($\tilde{R},\tilde{\rho}$) space. The outflow is accelerated from the surface of the MAD disc and then smoothly pass through two critical points, i.e., the slow magnetosonic point and the fast magnetosonic point. The Alfv\'en point is not a critical point based on current dimensionless definition in equation \eqref{eq:dimensionless_paras}, the Bernoulli equation does not show an explicit singularity at the Alfvén point, when the poloidal velocity of the outflow successfully exceed the slow and fast model velocities, it must be exceed through the Alfv\'en velocity. The poloidal Alfv\'enic Mach number is defined as $M_{\rm A}^2=v_{\rm p}^2/v_{\rm Ap}^2$, the poloidal Alfv\'en velocity $v_{\rm Ap}=B_{\rm p}/\sqrt{4\pi \rho}$, based on the definition of the dimensionless parameters in equation \eqref{eq:dimensionless_paras} one can easily have $M_{\rm A}^2=\tilde{\rho}^{-1}$. At the radius $R=R_{\rm A}$ and density $\rho =\rho_{\rm A}$, where the Mach
number $M_{\rm A}=1$ is the Alfv\'en point, another two critical points (saddle points) can be obtained by requiring the partial
derivatives of $\tilde{H}(\tilde{R},\tilde{\rho}) $ to be zero, i.e.,
\begin{equation}\label{eq:Bernoulli_partial}
    \left. \frac{\partial \tilde{H}(\tilde{R},\tilde{\rho})}{\partial \tilde{R}}\right|_{\rm s,f} = \left.\frac{\partial \tilde{H}(\tilde{R},\tilde{\rho})}{\partial \tilde{\rho}}\right|_{\rm s,f} = 0.
\end{equation}
where the subscripts ${\rm {s,f}}$ refer to the quantities at slow and fast points. Along the magnetic field, the value of the Bernoulli integral is a constant,
\begin{equation}\label{eq:Bernoulli_const} 
        \tilde{H}(\tilde{R}_{\rm s},\tilde{\rho}_{\rm s}) = \tilde{H}( \tilde{R}_{z_{\rm h}}, \tilde{\rho}_{z_{\rm h}})=\tilde{H}(\tilde{R}_{\rm f},\tilde{\rho}_{\rm f}).
\end{equation}
At the footpoint, fast and slow point, six equations are introduced, i.e., equations \eqref{eq:Bernoulli_partial} and \eqref{eq:Bernoulli_const} at $(\tilde{R}_{z_{\rm h}}, \tilde{\rho}_{z_{\rm h}})$, $(\tilde{R}_{\rm s},\tilde{\rho}_{\rm s})$ and $(\tilde{R}_{\rm f},\tilde{\rho}_{\rm f})$. Based on the given magnetic field configurations by equations \eqref{eq:Bz_midplane} and \eqref{eq:magnetic flux}, these six equations can be solved when the temperature and gas density at the surface of the MAD disc are specified as boundary conditions,
and then the dynamics of the outflow along the field line can be derived. The mass loss rate per unit area of the outflow is given by \cite[][]{2013ApJ...765..149C,2019ApJ...872..149L} 
\begin{equation}\label{eq:mass loss rate per unit area} 
    \dot{m}_{\rm{w}} = \kappa B_{\rm z}(R_{z_{\rm h}},z_{\rm h}) = \frac{\rho_{\rm s}v_{\rm p,\rm s}}{B_{\rm p,\rm s}}B_{\rm z}(R_{z_{\rm h}},z_{\rm h}).
\end{equation}
The total mass-loss rate in the outflows is
\begin{equation} \label{eq:total mass loss rate}
    \dot{M}_{\rm{w}} = \int_{R_{\rm{in}}}^{R_{\rm{out}}} 4\pi R \dot{m}_{\rm{w}} \, dR,
\end{equation}
where $ R_{\rm in}$ and $R_{\rm out}$ are the inner and outer edge of the disc, respectively. For given boundary conditions at the base of the outflow (i.e., temperature and density of the outflowing gas), the properties of the outflow are available with the calculations described in this section. Equation \eqref{eq:angular momentum conser} shows that the total specific angular momentum carried by the outflow launched from radius $R$ is
\begin{equation}
    l_{\rm w}(R)=\Omega(R)R^2_{\rm A}(R),
\end{equation}
the total angular momentum loss rate carried vertically by the outflow is consequently 
\begin{equation} \label{eq:outflow angular momentum}
   {\dot{J}_{\rm{w}} = \int_{R_{\rm{in}}}^{R_{\rm{out}}} 4\pi R \dot{m}_{\rm{w}}(R)\Omega(R) R_{\rm A}^2(R) \, dR,}
\end{equation}
for a field line anchored in the disc, the specific angular momentum of the disc material at its footpoint is
\begin{equation}
    l_{\rm d}=\Omega(R)R^2.
\end{equation}
It is useful to define a dimensionless magnetic lever arm parameter as
\begin{equation}\label{eq:angular momenten ratio}
    \tilde{l}=\frac{l_{\rm w}}{l_{\rm d}}=\left(\frac{R_{\rm A}}{R}  \right)^2,
\end{equation}
this parameter describes the efficiency on which the magnetically driven outflows extract angular momentum from the accretion disc. The total angular momentum carried by the outflow contains both the angular momentum originally associated with the local disc material and the additional angular momentum transferred magnetically from the disc into the outflow. The latter corresponds to the torque exerted by the outflow acting on the disc. A large value of $R_{\rm A}/R$, and hence of $\tilde{l}$, therefore allows a relatively small amount of outflowing material to extract a substantial amount of angular momentum from the disc.
For comparison, the net angular momentum flux entering the disc between its outer and inner boundaries may be roughly estimated as
\begin{equation} \label{eq:disc angular momentum}
    \dot{J}_{\rm{d}} = \dot{J}_{\rm{d,out}} - \dot{J}_{\rm{d,in}}=\dot{M}_{\rm acc}(R_{\rm out})R_{\rm out}^2\Omega(R_{\rm out}) - \dot{M}_{\rm acc}(R_{\rm in})R_{\rm in}^2\Omega(R_{\rm in}).
\end{equation}

\section{Numerical Setup}\label{sec:NuSet}
 We set the outer boundary of the MAD disc at $1000R_{\rm in}$, where $R_{\rm in}=6R_{\rm g}$ is the inner radius of the disc. In our calculations, the magnetic field footpoint radius extends from ${R_{\rm zh,min}=5R_{\rm in}}$ to ${1000R_{\rm in}}$. Thus, the outflow solutions considered in the present work are launched outside the strongly relativistic region close to the black hole.  We therefore expect general-relativistic corrections to the gravitational potential and the outflow-launching conditions to be modest over the radial range considered here. We consequently adopt the Newtonian gravitational potential in equation \eqref{eq:efftive potential} as a reasonable approximation for investigating the global magnetocentrifugal acceleration of the outflow. The present model is not intended to describe the strongly relativistic flow close to the event horizon or the innermost stable circular orbit. We note, however, that special-relativistic corrections associated with the outflow velocity itself may still become important for the fastest solutions, as discussed in Section \ref{sec:CandD}. The configurations of the large-scale magnetic field and the dynamics of the magnetically driven outflow have been established above. As discussed in Section \ref{sec:model}, the dynamical structure of the accretion flow, the large-scale field configurations, and the properties of the outflows are a tightly coupled system. 
To construct a fiducial model representative of a MAD, we adopt the radial structure measured from the high-resolution GRMHD simulations of \cite{2022ApJ...941...30C}. These simulations provide physically motivated profiles of the density, rotation, magnetization, and disc thickness appropriate for a saturated MAD. Based on the dynamical structure of their simulation (see Figures 8 and 15 in \cite{2022ApJ...941...30C}), we adopt following prescriptions as our fiducial disc structure,
\begin{equation}\label{eq:parameter former two}
   \rho(R,0) \propto R^{-1.1},
   \alpha(R)\propto R^{-0.4},
    f_{\Omega}(R)=\frac{\Omega}{\Omega _{\rm K}}\propto R^{0.07},
    \beta_{z_{\rm h}}(R,z_{\rm h}) \propto  R^{0.4}.
\end{equation}
We adopt the disc properties that were reported in \cite{2022ApJ...941...30C} as our fiducial model parameters, i.e., we set $\alpha(R_{\rm in})=0.3$, $f_\Omega(R_{\rm out})=0.8$ and  $\beta_{z_{\rm h}}(R_{\rm out})=1$. The footpoint of the magnetic field lines (also referred to as the base of the outflow, we further assume that the outflow base is at a certain disc scale height) is set to be at $H/R=0.3$ and the dimensionless temperature $\Theta_{z_{\rm h}}\sim (H/R)^2$. The gas density at the base of the outflow is evaluated as $\rho(R,z_{\rm h})=\epsilon \rho(R,0)$ and $\epsilon=10^{-3}$ is adopted as a fiducial value. The adiabatic index is set to $\gamma=1.4$ and we also calculated  $\gamma=1.5$ as a comparison. With above initial condition, the dimensionless density, Bernoulli constant, and the locations of the critical points related to a specific field line are available by solving a set of six equations (i.e., equations \eqref{eq:Bernoulli_partial} and \eqref{eq:Bernoulli_const}) numerically, the obtained outflow solution passes smoothly through the critical points. The dimensionless density of the outflow along the field line can be re-calculated with the Bernoulli equation \eqref{eq:Bernoulli dimensionless}. Then, we can calculate the other physical quantities (e.g., poloidal/azimuthal velocities, and temperature) of the outflow with equations \eqref{eq:induc_poloi}, \eqref{eq:angular momentum conser}, and \eqref{eq:mass_conser}. Repeating the calculations along each field line given by equation \eqref{eq:magnetic flux}, we obtain the global dynamics of the outflow.

\subsection{Comparison with Classical Magnetocentrifugal outflow Models}
The steady magnetohydrodynamic framework adopted in this work is based on the classical theory of magnetically driven outflow.  \cite{1967ApJ...148..217W} established the basic steady MHD description of a rotating magnetized outflow and demonstrated how magnetic stresses extract angular momentum from the rotating central object. \cite{1985A&A...152..121S} extended this type of treatment to two-dimensional, axisymmetric magnetized outflows. In particular, \cite{1994A&A...287...80C},  \cite{1995ApJ...452L..41K} and \cite{1997ApJ...474..362K} investigated steady, axisymmetric magnetically driven outflows along prescribed poloidal magnetic field anchored in a Keplerian thin accretion disc, and further studied their mass loading, acceleration, and terminal velocities, which are closely related to the approaches we adopted.

In this work, we apply the classical trans-magnetosonic outflow formalism to a physically distinct MAD regime, with the disc boundary conditions calibrated by GRMHD simulations, and investigate how the strong magnetization, sub-Keplerian rotation, and geometrical thickness characteristic of MADs modify the outflow acceleration, critical surfaces, mass lost, and terminal velocity. We focus on investigating the outflow solutions under conditions of a MAD. The dynamics of the outflow-base are constrained by the radial structure found in numerical simulations of saturated MADs, which are geometrically thick, rotate sub-Keplerianly, and are threaded by dynamically important large-scale magnetic fields. In our fiducial model, the disc thickness is ($H/R\simeq0.3$), the field-line angular velocity is sub-Keplerian, ($f_{\Omega}\equiv\Omega/\Omega_{\rm K}<1$), and the magnetic pressure is comparable to or exceeds the gas pressure near the outflow base, ($\beta_{\rm zh}\lesssim1$). Moreover, the density, rotation, and magnetization vary systematically with radius according to the MAD structure adopted from the simulations. Thus, the MAD model considered here represents a particular combination of strong magnetization, sub-Keplerian rotation, and radially varying outflow-base dynamical properties. This work aimed to explore the classical trans-magnetosonic solutions in this MAD-specific parameter regime. We focus on how these conditions determine the mass outflowing, the locations of the slow magnetosonic, Alfvén, and fast magnetosonic points, and the acceleration of the fast outflows, we further evaluate the total mass and angular-momentum fluxes carried by the outflows.

\section{Results} \label{sec:results}
\begin{figure}
    \centering
    \includegraphics[width=\linewidth]{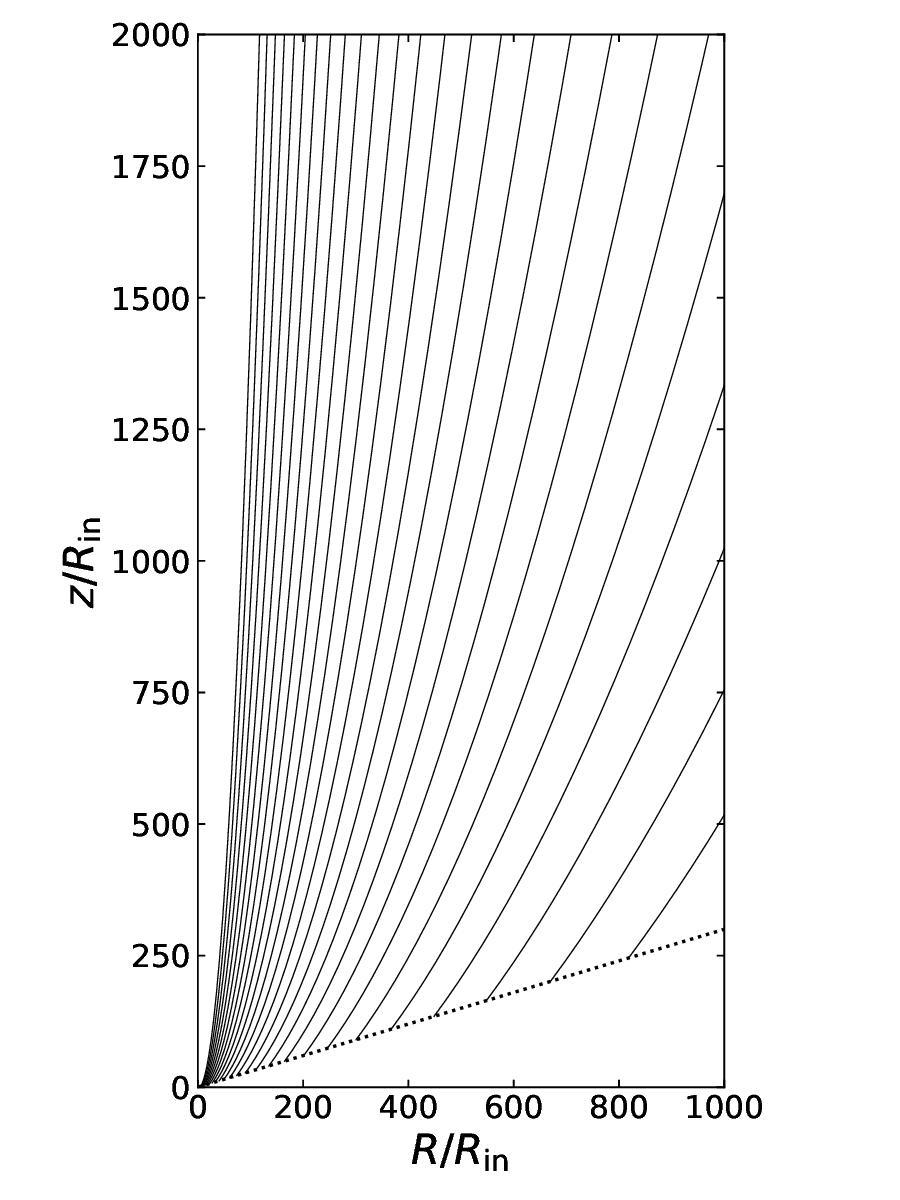} 
    \caption{The large-scale poloidal magnetic field configuration (solid curves) derived from equation \eqref{eq:magnetic flux}, the dotted line corresponds to the half disc thickness $H/R =0.3$.}
    \label{fig:magnetic line}
\end{figure}

\begin{figure}
    \centering
    \includegraphics[width=\linewidth]{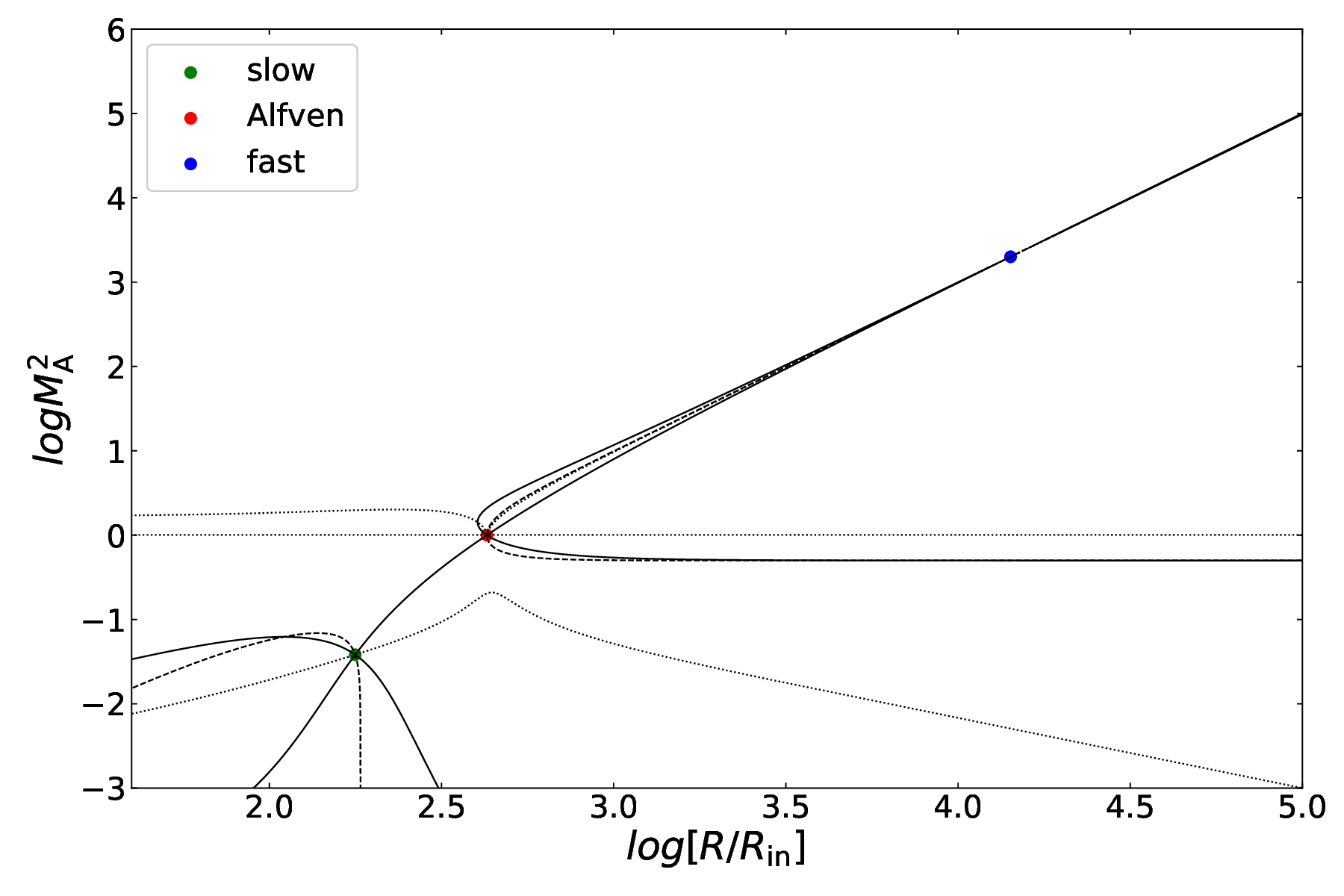}
    \caption{Solution plane of the magnetically driven outflow. $M_{\rm A}$ is the poloidal Alfv\'enic Mach number, which is defined as the ratio of poloidal velocity to poloidal velocity at Alfv\'en point, i.e., $M_{\rm A}^2={v_{\rm p}^2}/{v_{\rm Ap}^2}$. The solid, dashed and dotted curves represent the solution of equations $\tilde{H}(\tilde{R},\tilde{\rho})=\tilde{E}, {\partial \tilde{H}}/{\partial \tilde{R}}= 0, {\partial \tilde{H}}/{\partial \tilde{\rho}}= 0$, respectively. The intersection points of the curves from the lower left to the upper right are the slow sonic point (green), the Alfv\'en point (red), and the fast sonic point (blue).}
    \label{fig:contour_MA_gamma_1.4}
\end{figure}

\begin{figure*}
\begin{center}
     \includegraphics[width=0.6\textwidth]{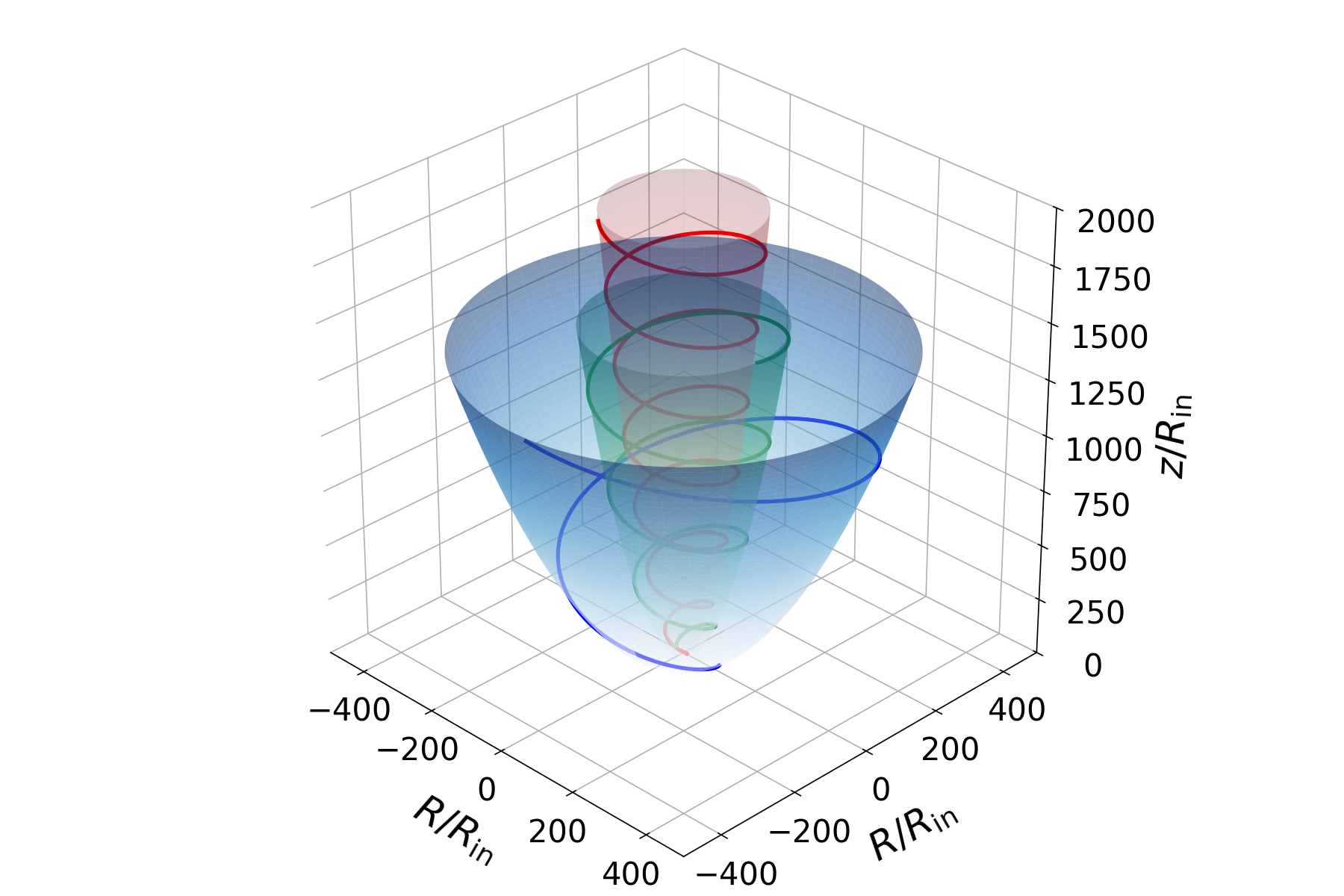}
     \includegraphics[width=0.35\textwidth]{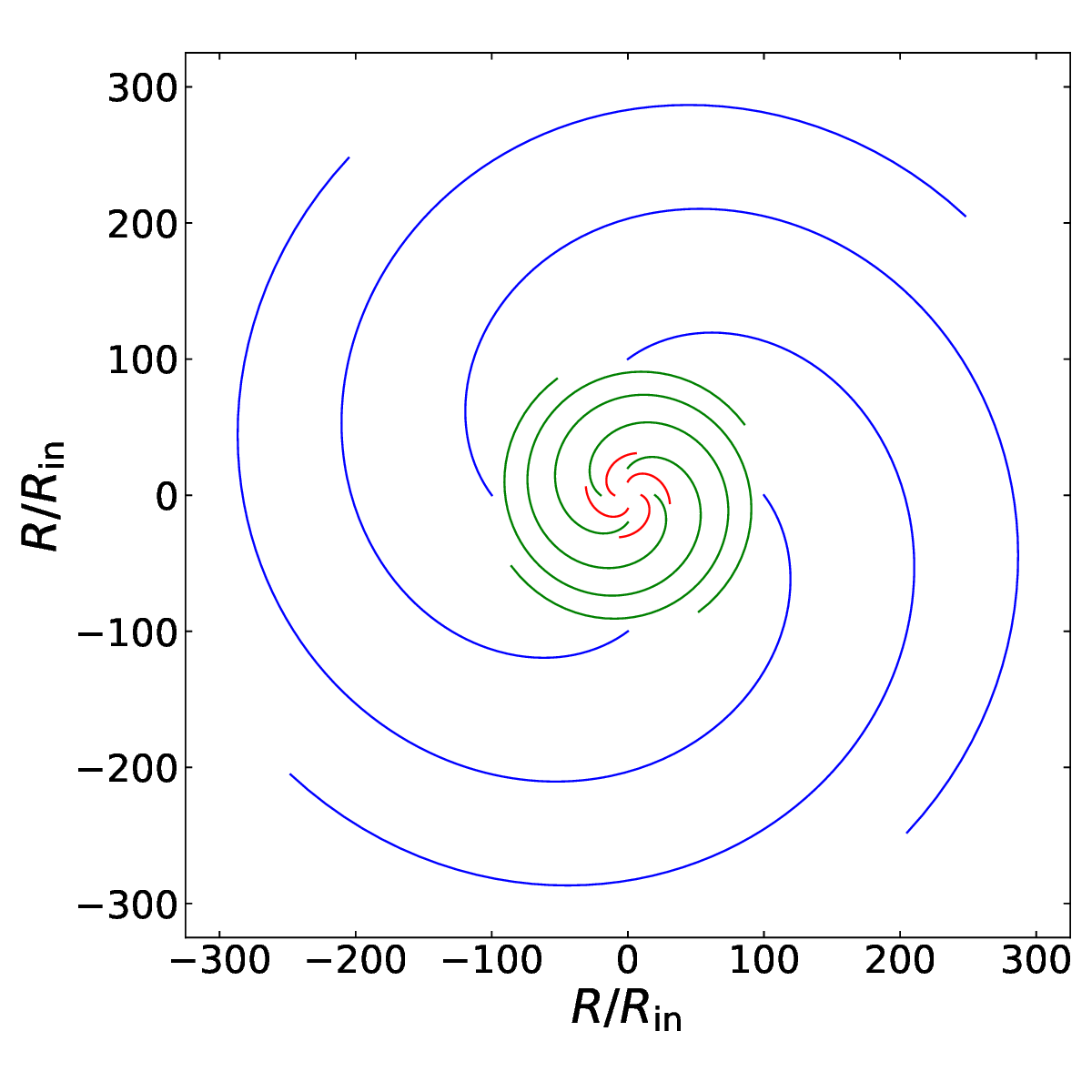}     
\end{center}
     \caption{Configurations of the field lines. Left panel: magnetic field lines and their spatial distribution. Right panel: the corresponding projections onto the disc midplane, the footpoint of the magnetic field lines are at $R_{z_{\rm h}}/R_{\rm in}=10 ({\rm red}), 20 ({\rm green}), 100 ({\rm blue})$, respectively, the two adjacent lines have a toroidal angle difference of $\pi/2$.} 
    \label{fig:magneticFieldLine spatial distribution and projection}
\end{figure*}

\begin{figure*}
     \includegraphics[width=0.45\textwidth]{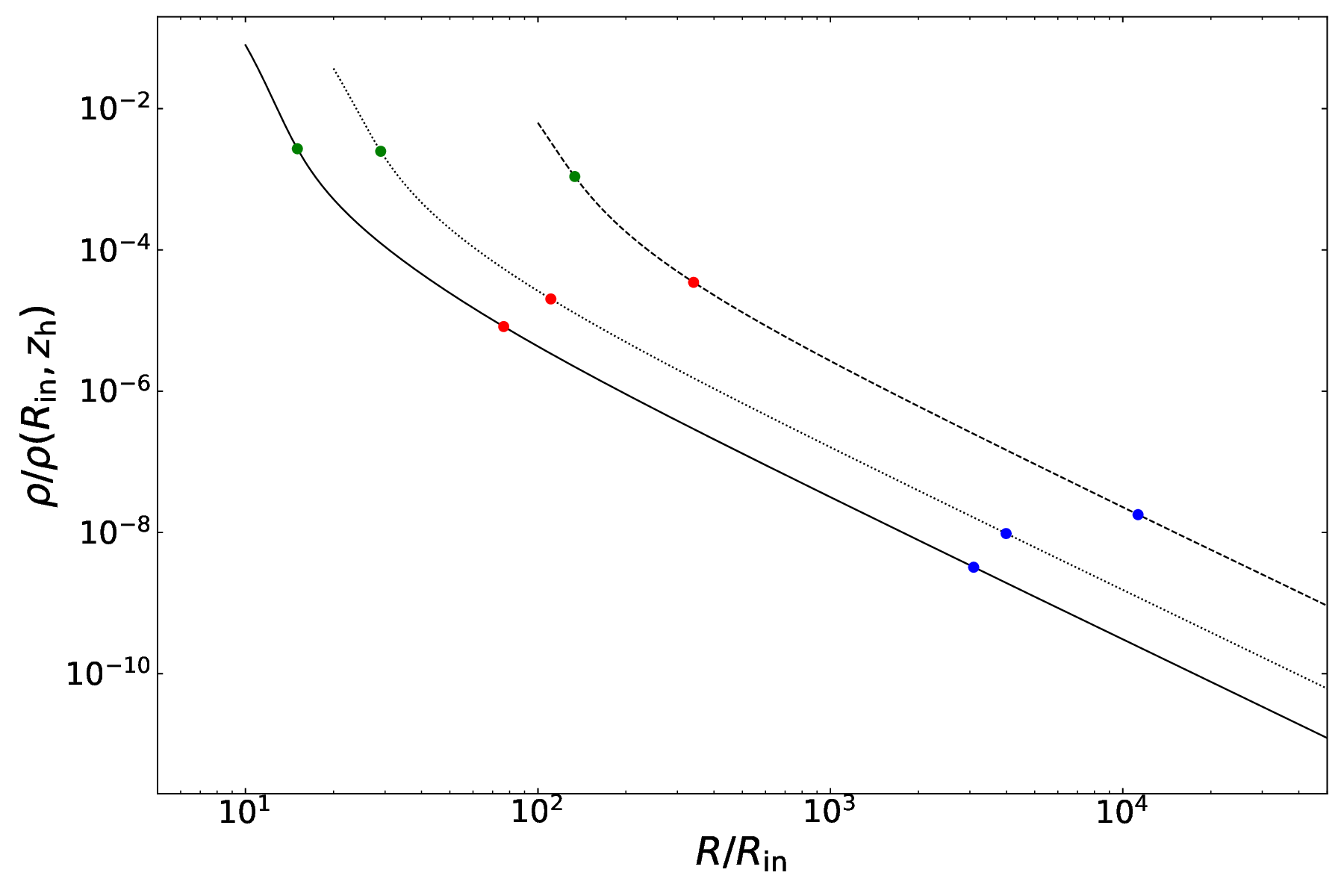}
     \includegraphics[width=0.45\textwidth]{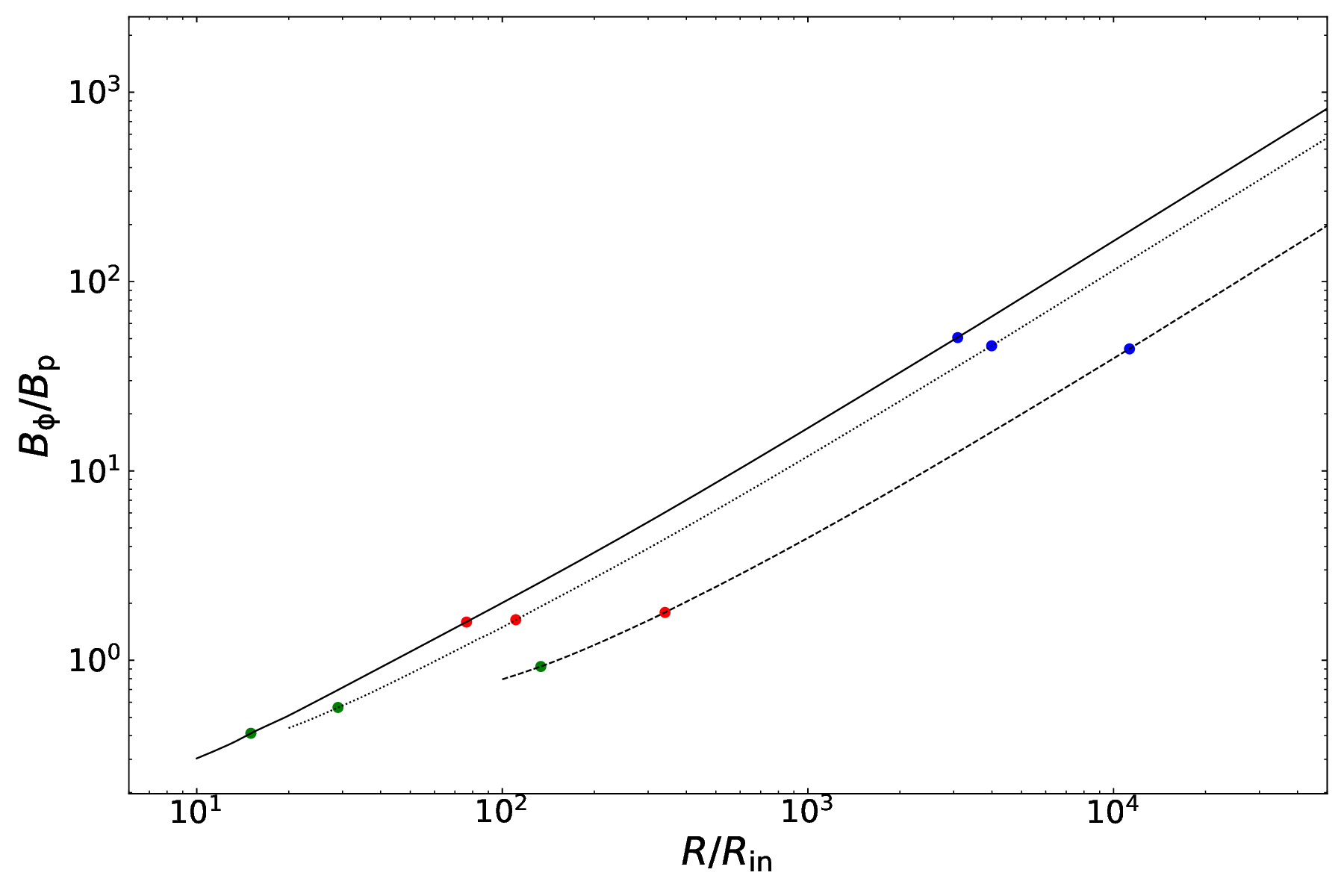} 
     \includegraphics[width=0.45\textwidth]{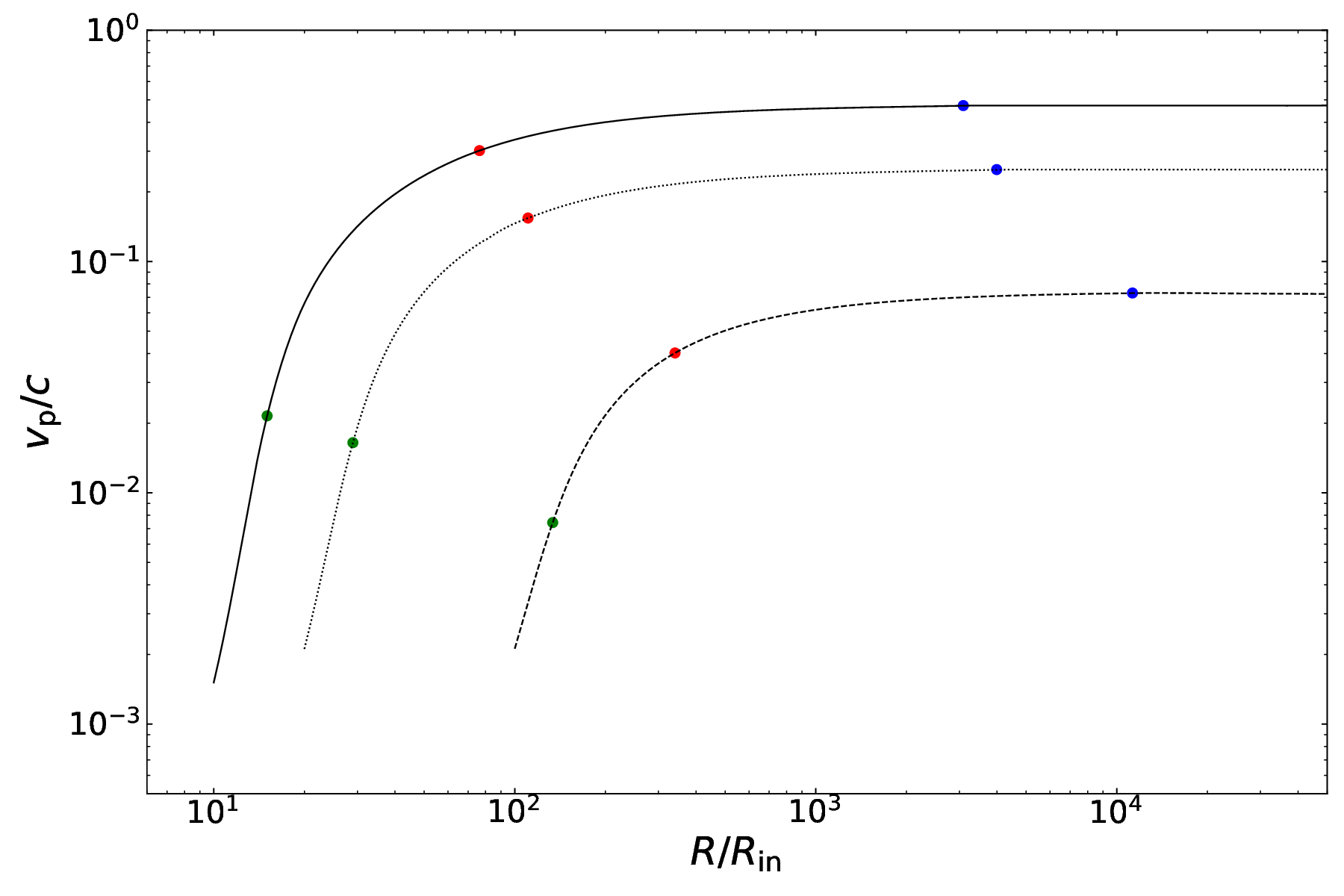}
     \includegraphics[width=0.45\textwidth]{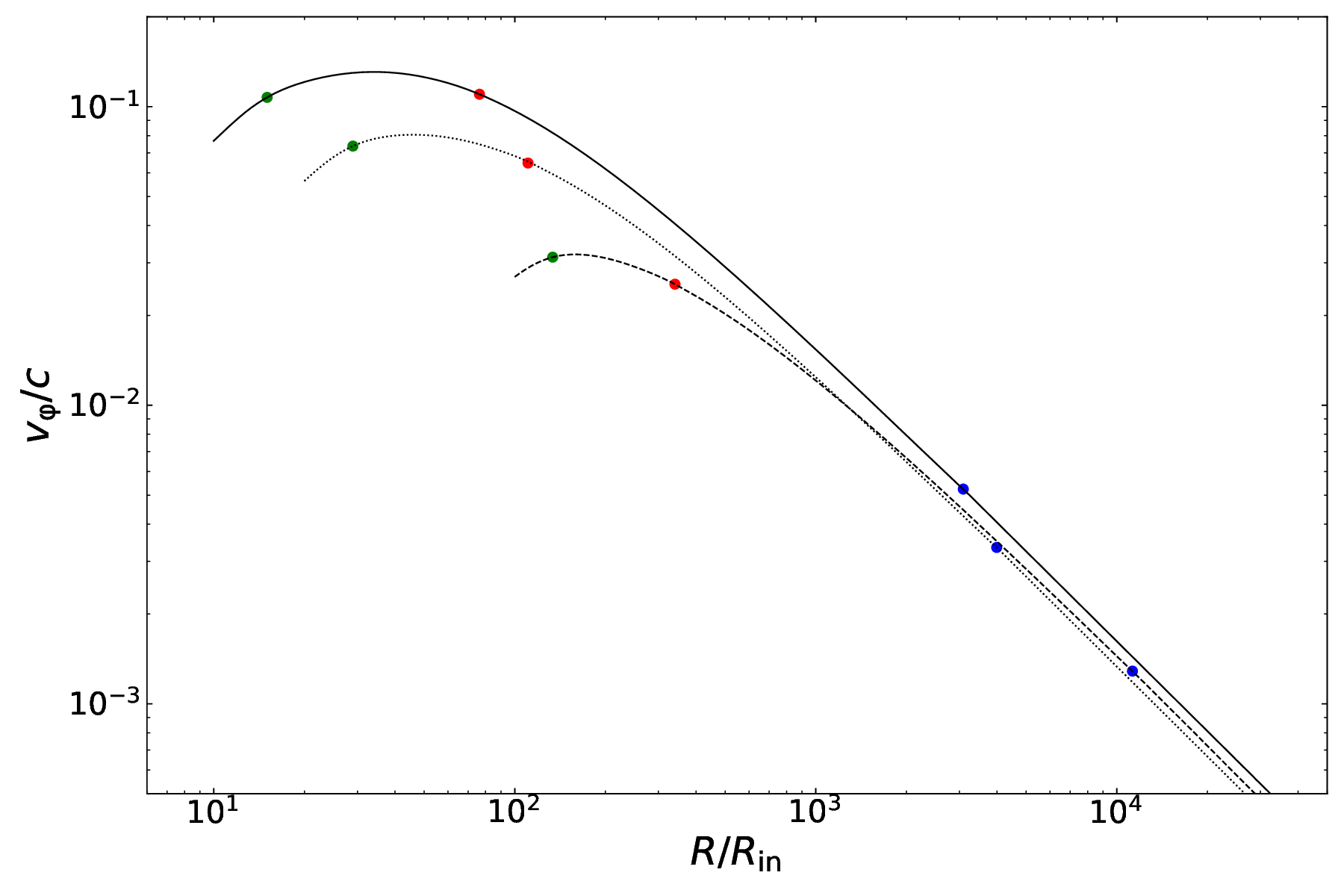}
     
 \caption{The variables of density, magnetic field components, poloidal and toroidal velocities varies with the magnetic field line. The radius of the field line footpoint are $R_{z_{\rm h}}=10R_{\rm in}\; {\rm (solid)}, 20R_{\rm in}\; {\rm (dotted)}, {\rm and}\; 100R_{\rm in}\;{\rm (dashed)}$. The points with different colors are related to the slow sonic point (green),  Alfv\'en point (red), and fast sonic point (blue).}
	\label{fig:variables}
\end{figure*}

The configurations of the large-scale magnetic field, shown in Figure \ref{fig:magnetic line}, are derived from equation \eqref{eq:magnetic flux}. As discussed in Section \ref{sec:model}, the footpoint of the field line is set at $H/R\sim 0.3$ (see the dotted line in Figure \ref{fig:magnetic line}). Figure \ref{fig:contour_MA_gamma_1.4} shows the solution plane, ($M_{\rm A}^2, R$), of the outflow equations, the green/blue and red dots are related to the slow/fast magnetosonic point (saddle point) and Alfvén point, the solution curve smoothly connect these three points as the poloidal velocity of the outflow were accelerated up to the slow-, Alfvén-, and fast-mode velocities. The spatial structures of the field lines and their projection onto the equatorial plane are plotted in Figure \ref{fig:magneticFieldLine spatial distribution and projection}. The variables of density, magnetic field components, poloidal and toroidal velocities varies along the magnetic field line are plotted in Figure \ref{fig:variables}. The outflow is centrifugally accelerated along the magnetic field lines, beyond the Alfv\'en point, as the poloidal velocity increase the density and the toroidal velocity of the outflow decrease simultaneously. Within the   Alfv\'en point, the poloidal component of the magnetic field dominates over the toroidal component, since most of the acceleration (see the fast increase of the poloidal velocity between the footpoint and Alfv\'en point) then takes place in the magnetically dominated region roughly within the Alfv\'en point. Beyond the Alfv\'en point the co-rotation of the outflowing gas with the field line fails, a relatively strong wound up field develops and the toroidal component gradually dominates the poloidal component. We also calculated the related variables with $\gamma =1.5$ for a comparison, see Figure \ref{fig:variablesGamma1p5}.

\begin{figure*}
     \includegraphics[width=0.7\textwidth]{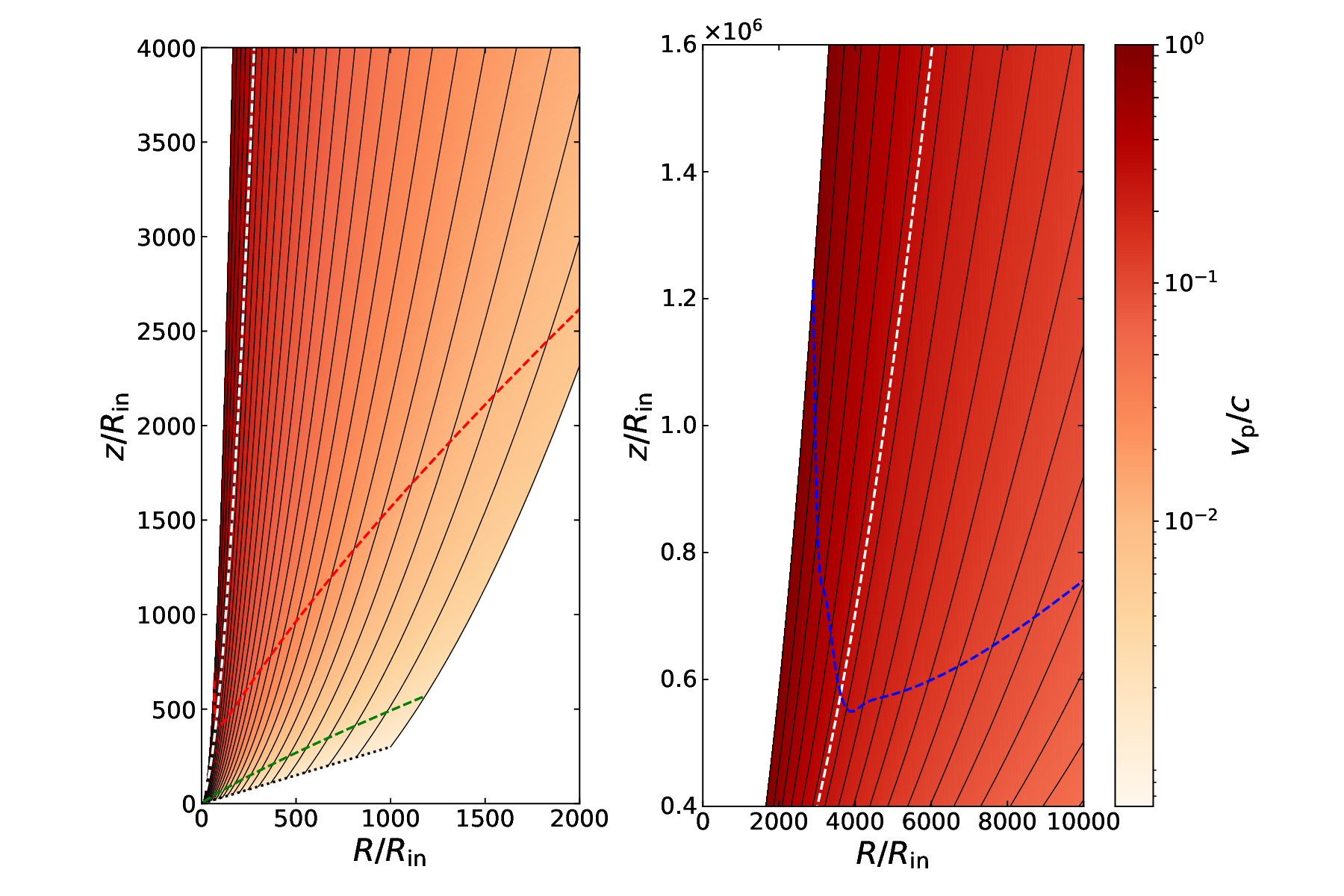}
     \includegraphics[width=0.7\textwidth]{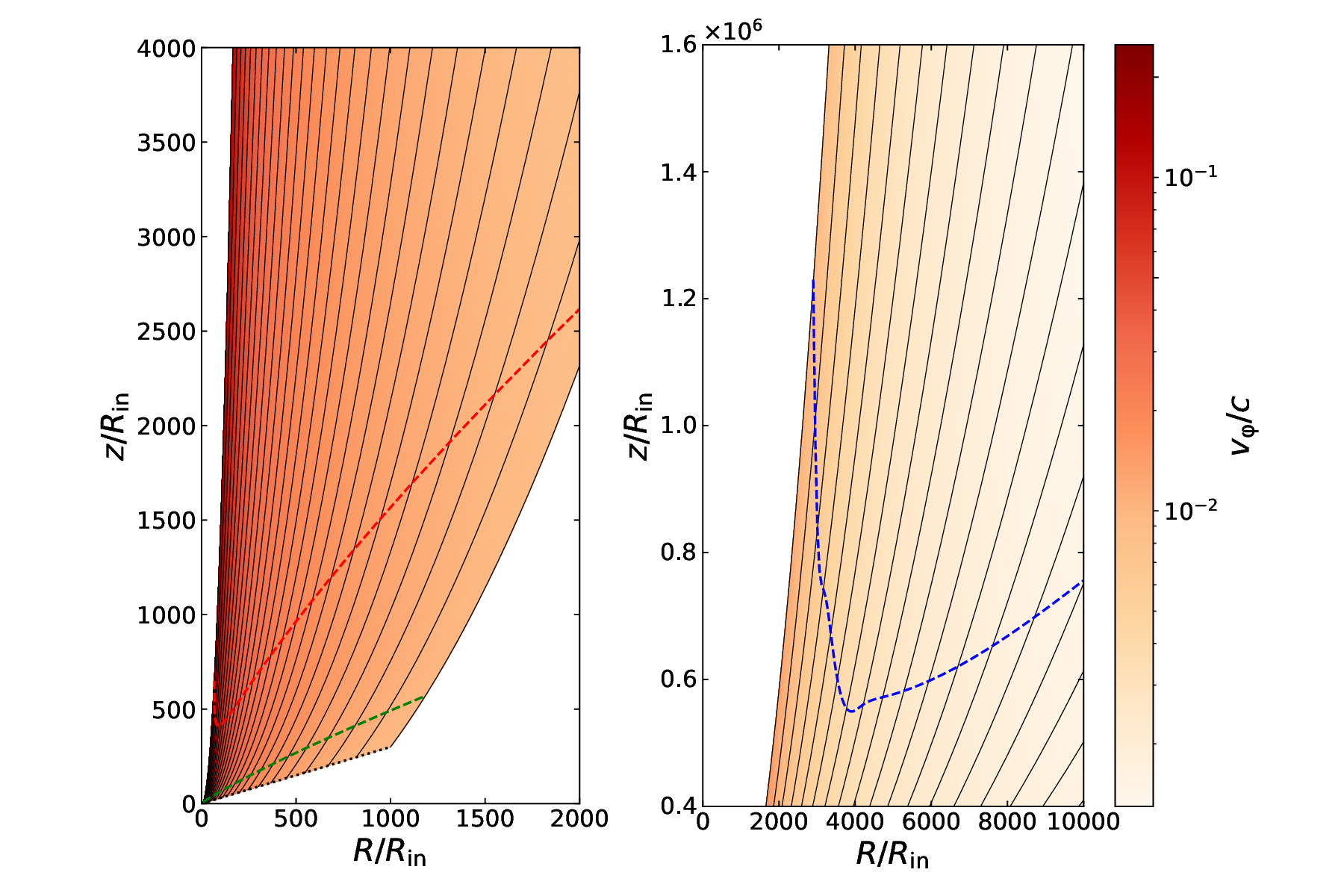} 

 \caption{Velocity profiles of the magnetically driven outflows (because of the large radial distance of the fast sonic points, we use two diagrams to illustrate the velocity profile, ensuring that critical point locations are displayed). Upper panel: the poloidal velocities ($v_{\rm p}$) of the outflows. Lower panel: the toroidal velocities ($v_{\rm \varphi}$) of the outflows. Both of the velocity profiles are scaled by the light speed ($c$). The same as Figure \ref{fig:magnetic line}, the solid curves and dotted line represent the magnetic field lines and disc surface, respectively. The dashed lines represent the locations of the slow sonic point (green), Alfv\'en point (red) and fast sonic point (blue). The white dashed line in the upper panel is labeled for the magnetic field line, along which the speed of outflow is accelerated up to $0.3c$, outside this field line the velocity of the outflow is less than $0.3c$}
    \label{fig:vp_vphi_tripcolor_gamma_1.4}
\end{figure*}

\begin{figure*}
     \includegraphics[width=0.6\textwidth]{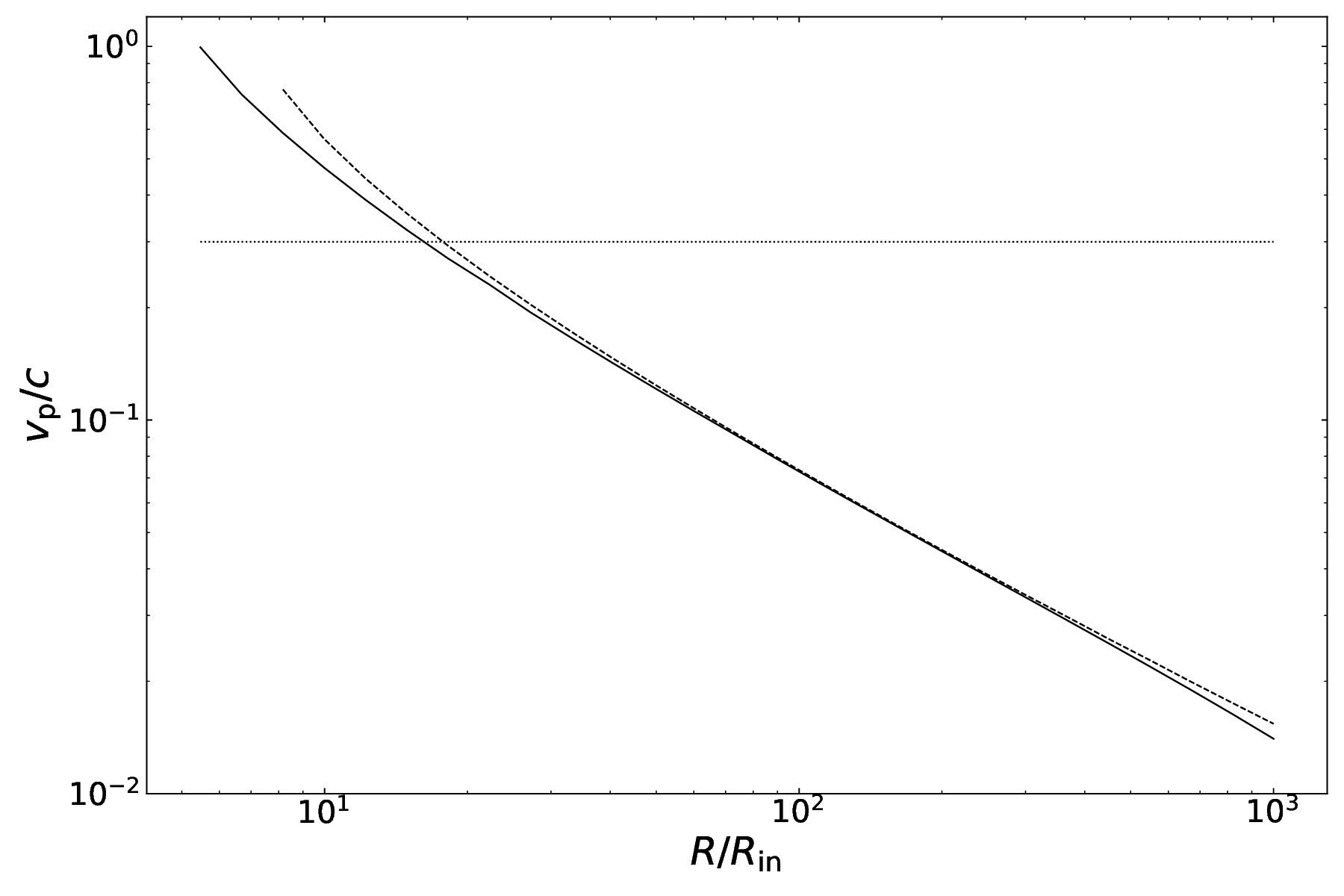}
 \caption{The velocity of the outflow at fast sonic point, the solid line is calculated for  $\gamma=1.4$ while dashed line is for  $\gamma=1.5$, the horizontal dotted line is labeled for the poloidal velocity equal to $0.3c$. }
    \label{fig:vp_fastpoint} 
\end{figure*}

\begin{figure*}
     \includegraphics[width=0.35\textwidth]{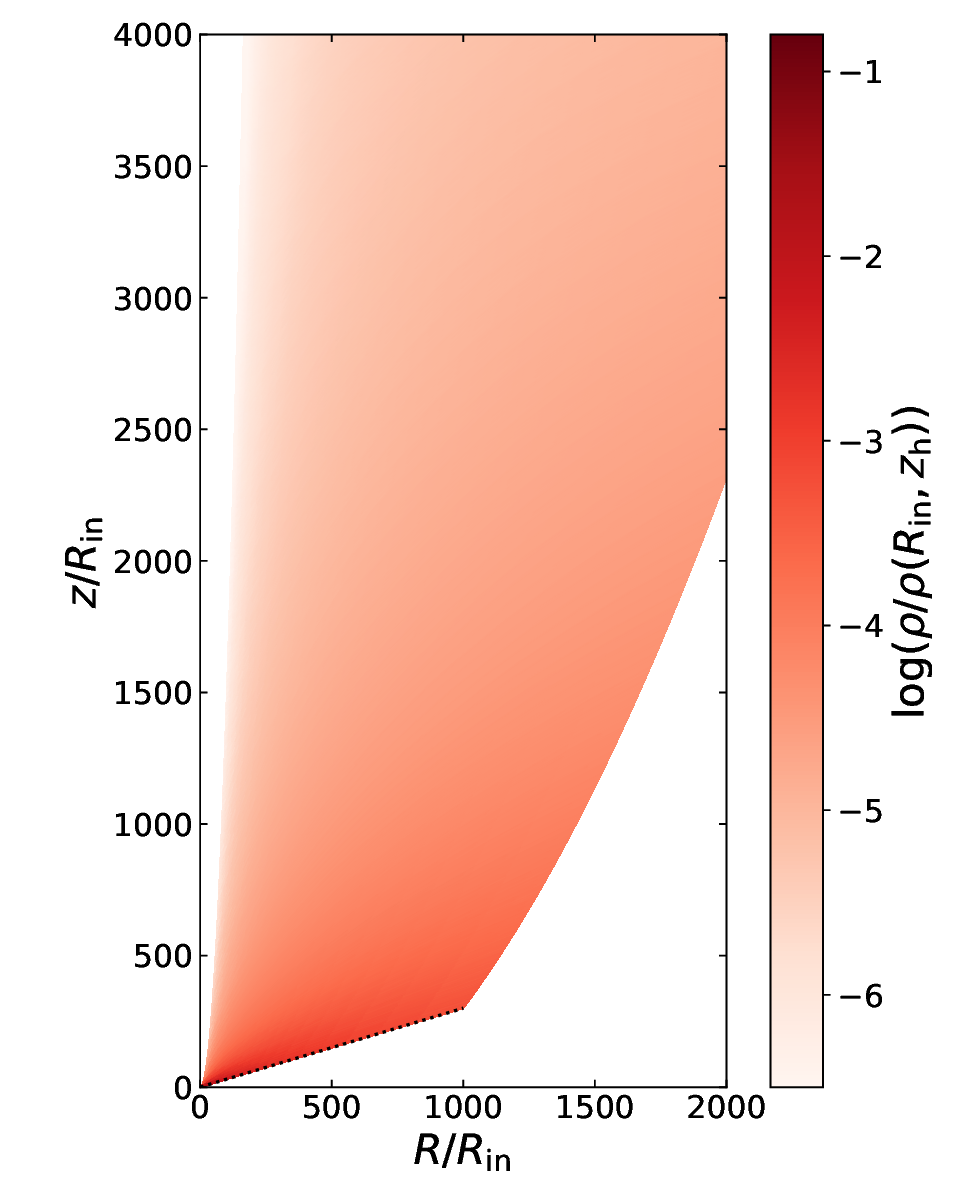}
     \includegraphics[width=0.35\textwidth]{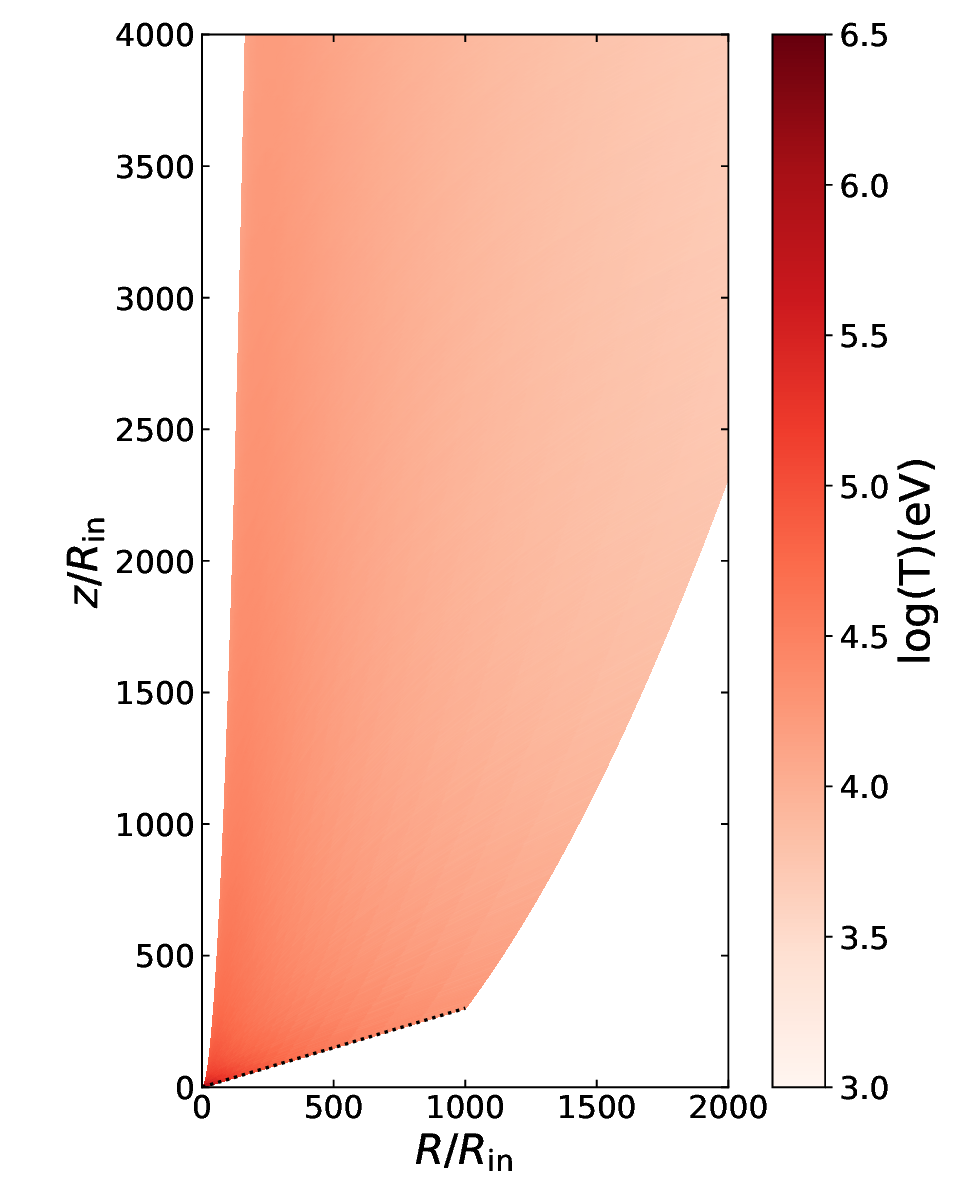} 

     \caption{Left panel: the density profile of the outflows, $\rho(R_{\rm{in}},z_{\rm{h}})$ is the gas density at the disc surface and radius $R=R_{\rm in}$. Right panel: the gas temperature profile of the outflows.}  
    \label{fig:Density and temperature profiles_gamma_1.4}
\end{figure*}

\begin{figure*}
     \includegraphics[width=0.45\textwidth]{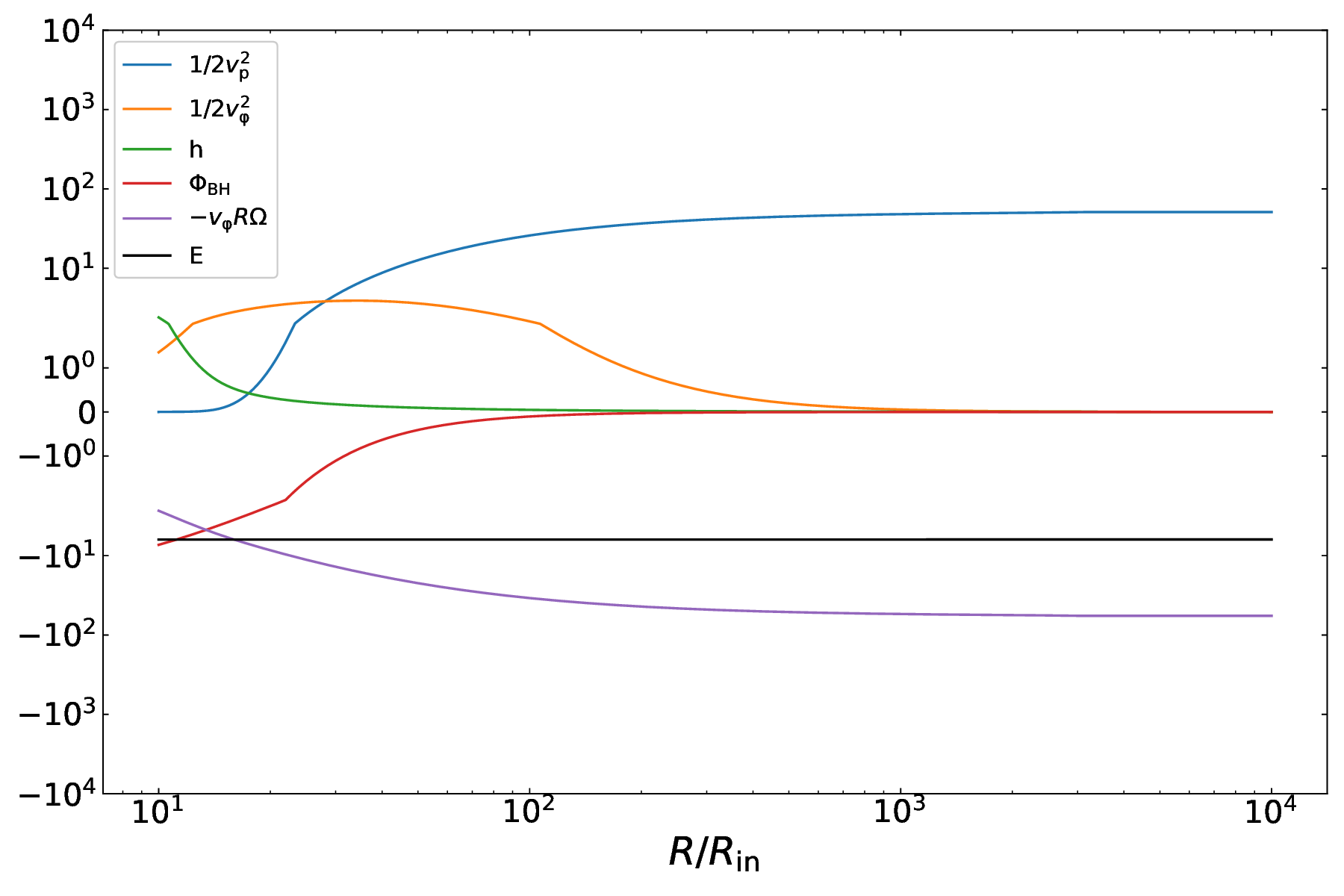}
     \includegraphics[width=0.45\textwidth]{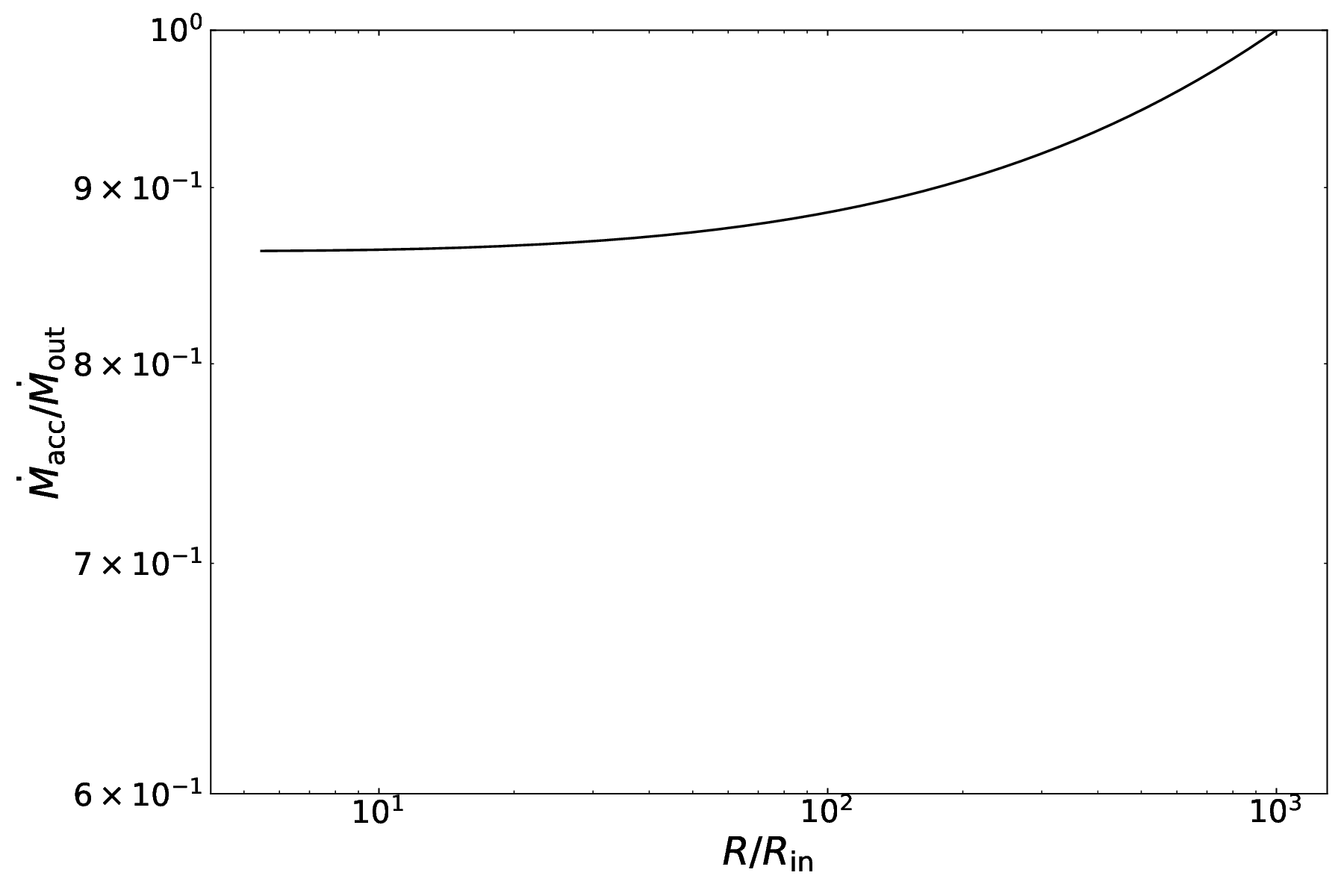} 

      \caption{Left panel: potential energies and Bernoulli constant as functions of the radius along a single magnetic line $(R_{z_{\rm h}}/R_{\rm in}=10)$. All the potential energies and Bernoulli constant are normalized by ${GM_{\rm BH}}/{R_{\rm A}}$. Right panel: the mass accretion rate as a function of radius. The $\dot{M}_{\rm out}$ is the mass accretion rate at the disc outer boundary.}
    \label{fig:Bernoulli_conservation and mass accretion rate_gamma_1.4}
\end{figure*}

The poloidal and toroidal velocities of the outflow region are plotted in Figure \ref{fig:vp_vphi_tripcolor_gamma_1.4}. Due to the strong magnetic field threading the inner MAD disc, the fast magnetosonic surface can be located far from the disc surface (see the blue dashed lines in Figure \ref{fig:vp_vphi_tripcolor_gamma_1.4}), allowing substantial magnetic acceleration along the field lines. The poloidal velocity increases continuously from the base of the outflow toward the fast magnetosonic surface, whereas the toroidal velocity reaches its maximum around the Alfv\'en radius and subsequently decreases as the flow propagates outward. For the outflow accelerated along inner field lines, the formal Newtonian solutions can reach velocities well above $0.3c$, and in the most strongly accelerated cases the calculated value approaches the mildly relativistic regime. The white dashed line in the upper panel of Figure  \ref{fig:vp_vphi_tripcolor_gamma_1.4} is labeled for the magnetic field line, along which the speed of outflow is accelerated up to $0.3c$. Within this field line, the velocity of formal Newtonian solution larger than $0.3c$, such high velocities should not be regarded as precise quantitative predictions because special-relativistic corrections become increasingly important as the velocity of the outflow increases. Instead, these solutions indicate that the strong large-scale magnetic field characteristic of the MAD regime provides sufficient magnetic energy to accelerate the outflow beyond the ordinary non-relativistic regime. In the velocity range ($v_p\lesssim0.3c$), where the Newtonian approximation remains considerably better controlled, the solutions naturally reach UFO-like velocities before entering the regime in which relativistic corrections become important. Also, in Figure \ref{fig:vp_fastpoint} we show the velocity of the outflow at fast sonic point, under current the formal Newtonian solution, the velocity of the outflow launched from $R\lesssim 20R_{\rm in}$ can reach above $0.3c$, outside this region the outflow is remain modest to middle relativistic speed. The outflow also exhibits a clear radial velocity stratification, field lines anchored at smaller disc radii generally produce faster outflows, whereas those launched farther out produce slower components. This is well consistent with the observation of multiple outflow components with different velocities originating from different radius in some AGNs or XRBs.

The densities of the outflow region are plotted in the left panel of Figure \ref{fig:Density and temperature profiles_gamma_1.4}. We adopted a polytropic equation of state with $\gamma =1.4$ the same as that adopted in the numerical simulation work done by \citep[][]{2018ApJ...857...34Z}, and further calculated the temperatures of the outflow (see right panel of Figure \ref{fig:Density and temperature profiles_gamma_1.4}), the temperature of the outflow is relatively high and roughly about several ten to hundred $\rm keV$ at large distance from the disc surface. But we note here that the specific value of the temperature is closely related to the heating and cooling process as the gas flowing outward, which is out of the scope of this work. We also calculated the related variables with $\gamma =1.5 $ for a comparison, and found that the results are not sensitive to the value of $\gamma$ (see Figure \ref{fig:Density and temperature profiles_gamma_1.5}). The locations of the slow sonic points and Alfv\'en points almost remain unchanged (see the green-dashed and red-dashed lines in the left panel of Figure  \ref{fig:vp_vphi_tripcolor_gamma_1.4} and \ref{fig:vp_vphi_tripcolor_gamma_1.5})

In the left panel of Figure \ref{fig:Bernoulli_conservation and mass accretion rate_gamma_1.4}, we plot the components of the Bernoulli equation, i.e., equation \eqref{eq:Bernoulli integral}, as the outflow accelerated along the magnetic field line, the poloidal kinetic energy ($\frac{1}{2}v_{\rm p}^2$) and gravitational potential ($\Phi_{\rm BH}=-GM_{\rm BH}/{(R^2+z^2)^{\frac{1}{2}}}$) gradually increased, while the enthalpy ($h$), rotational energy ($\frac{1}{2}v_{\varphi}^2$), and centrifugal potential  ($-v_{\varphi}R\Omega$) components decreased. In the right panel of  Figure \ref{fig:Bernoulli_conservation and mass accretion rate_gamma_1.4} we plot the mass accretion rate, only about $14\%$  of the accreting gas has been accelerated into the outflows, the relevant results calculated for $\gamma =1.5$ are plotted in Figure \ref{fig:Bernoulli_conservation and mass accretion rate_gamma_1.5}. The dimensionless magnetic lever arm in equation \eqref{eq:angular momenten ratio} is plotted in Figure \ref{fig:tilde_l}. For our fiducial MAD solutions, we find that $\tilde{l}$ is typically of order several tens and can reach values of order $10^2$ for some inner field lines. These values correspond to $R_{\rm A}/R\sim5-10$, indicating that the outflow remains magnetically coupled to the rotating disc over a large radial lever arm before crossing the Alfv\'en surface. The dynamically important large-scale poloidal field, together with the relatively low mass loading at the outflow-base, allows the magnetic field to enforce the corotation of the outflow to large distances from the disc, thereby moving the Alfvén surface outward. The resulting long magnetic lever arm is therefore a direct dynamical manifestation of the strongly magnetized MAD environment. Classical disc-outflow theory already predicts magnetocentrifugal acceleration and angular-momentum extraction, but the  Alfv\'en surface is relatively close to the disc surface \citep[][]{1994A&A...287...80C,2014ApJ...783...51C,2018ApJ...857...34Z,2022ApJ...926...11L}, but in the MAD parameter regime explored here the combination of strong magnetization and low mass loading result in the solution toward a large magnetic lever arms. We further evaluate how efficient the disc angular momentum removed by the magnetic outflow based on equations \eqref{eq:outflow angular momentum} and \eqref{eq:disc angular momentum}, it is found that about $55\%$ of the disc angular momentum is carried away by about $14\%$ mass loss in the outflow.

\begin{figure*}
     \includegraphics[width=0.45\textwidth]{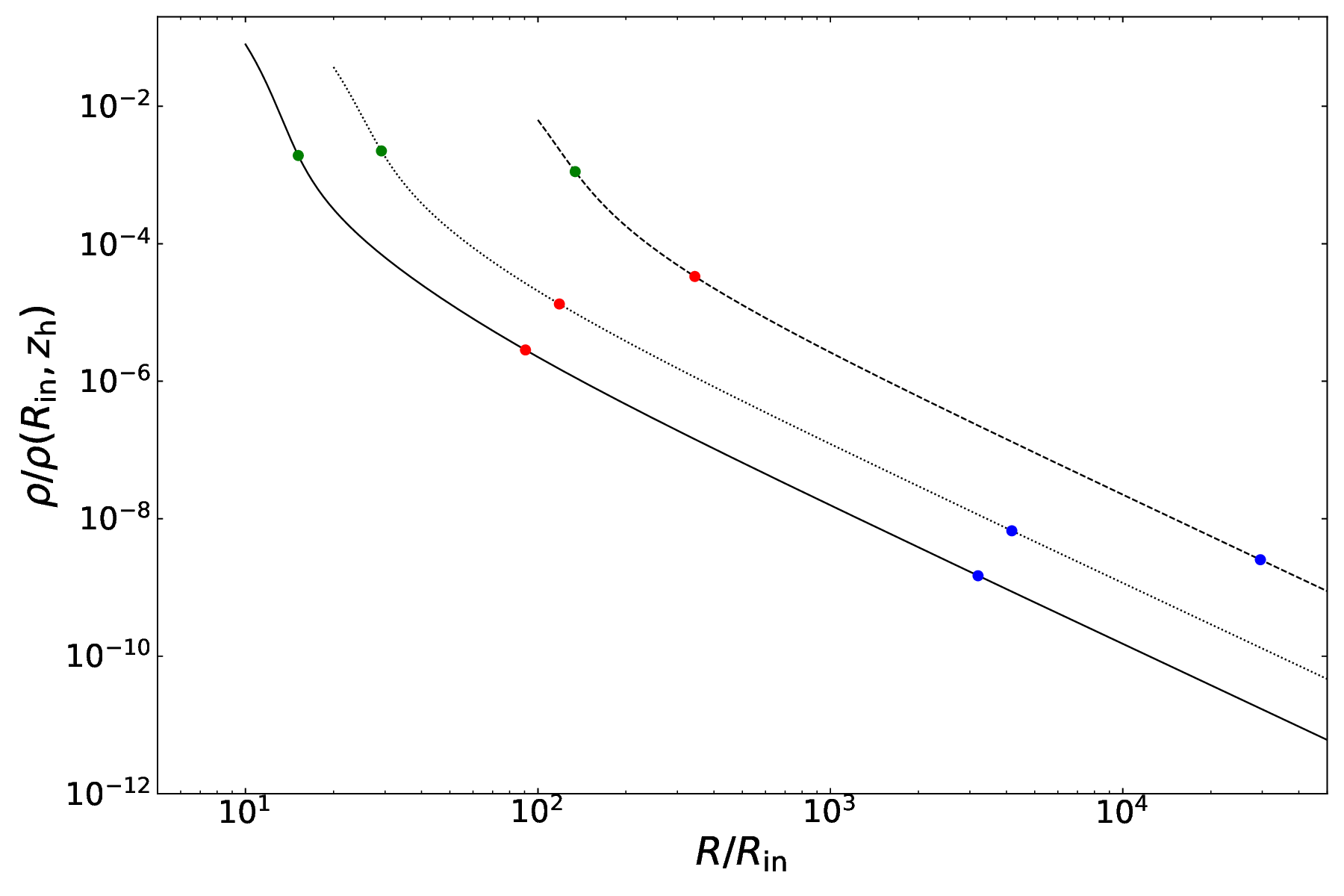}
     \includegraphics[width=0.45\textwidth]{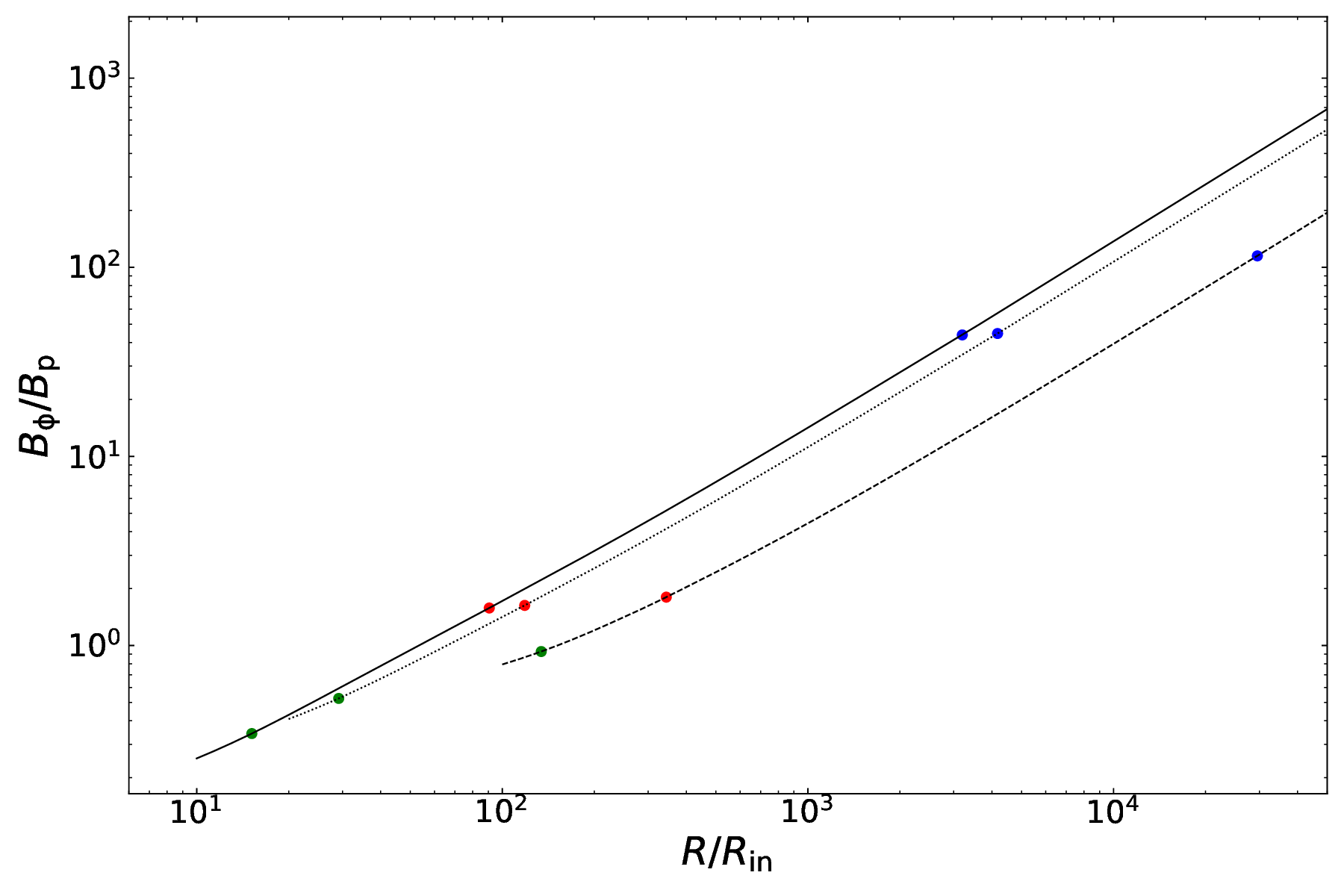} 
     \includegraphics[width=0.45\textwidth]{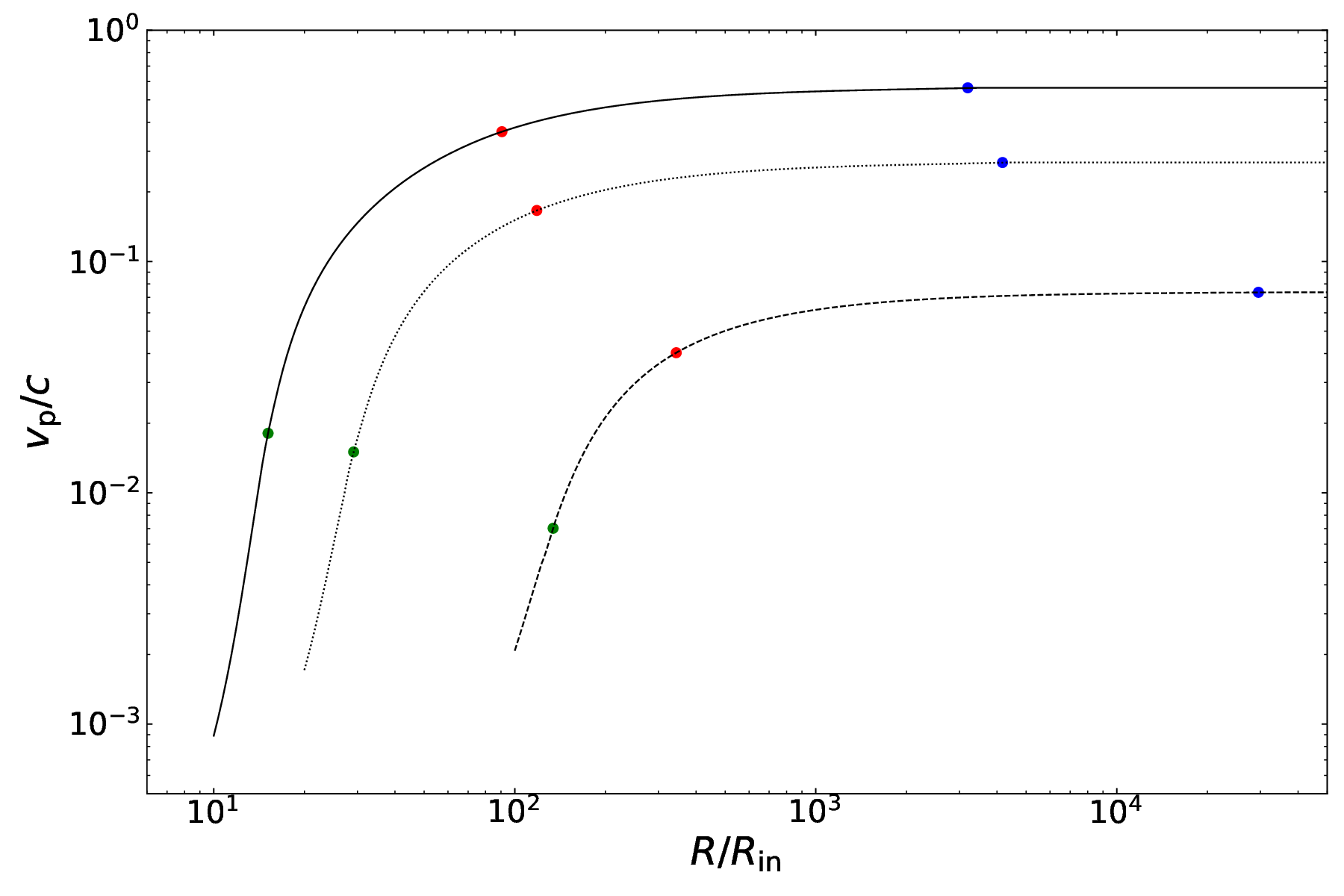} 
     \includegraphics[width=0.45\textwidth]{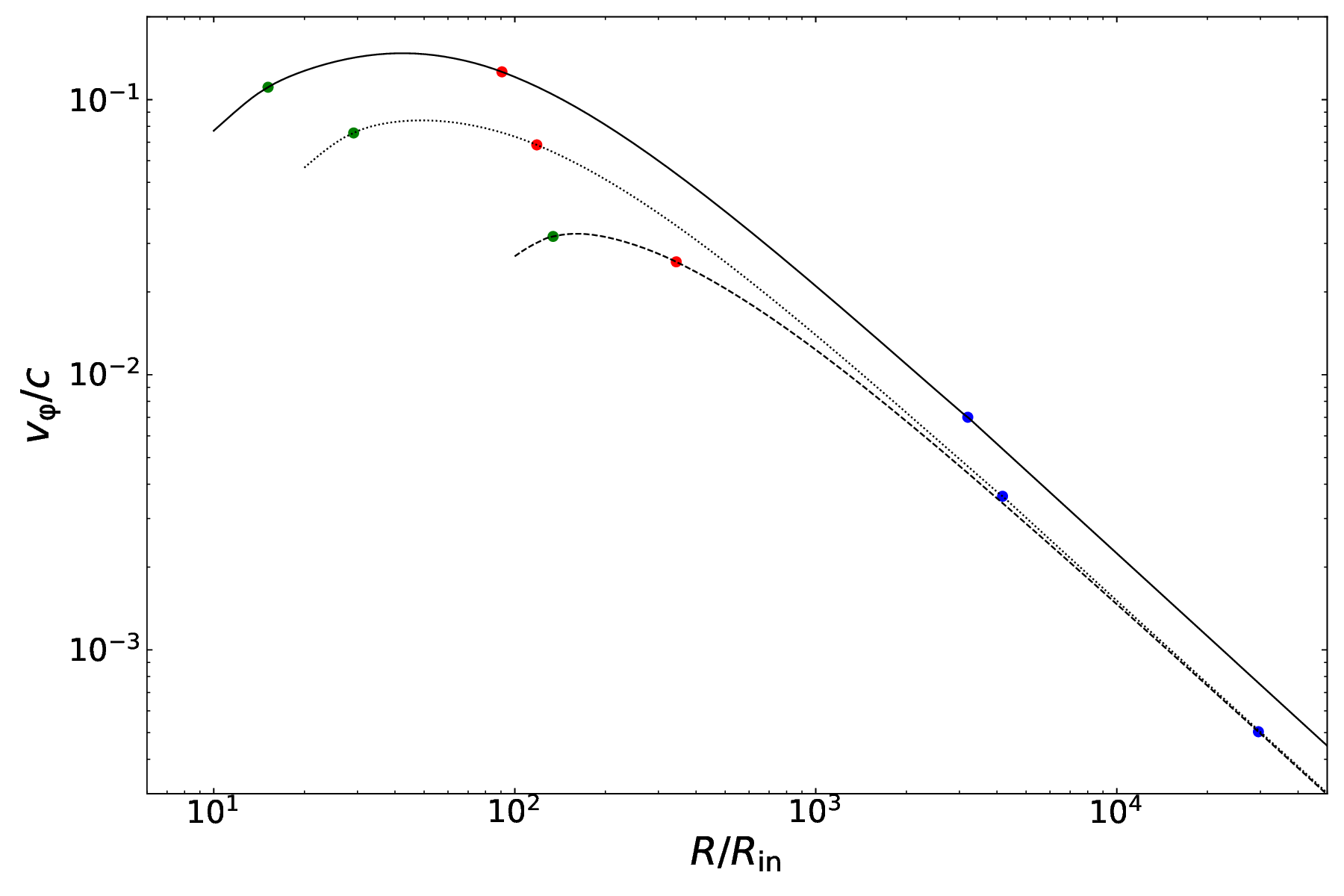} 

     \caption{The same as Figure \ref{fig:variables}, but calculated for $\gamma=1.5$}
	\label{fig:variablesGamma1p5}
\end{figure*}

\begin{figure*}
     \includegraphics[width=0.7\textwidth]{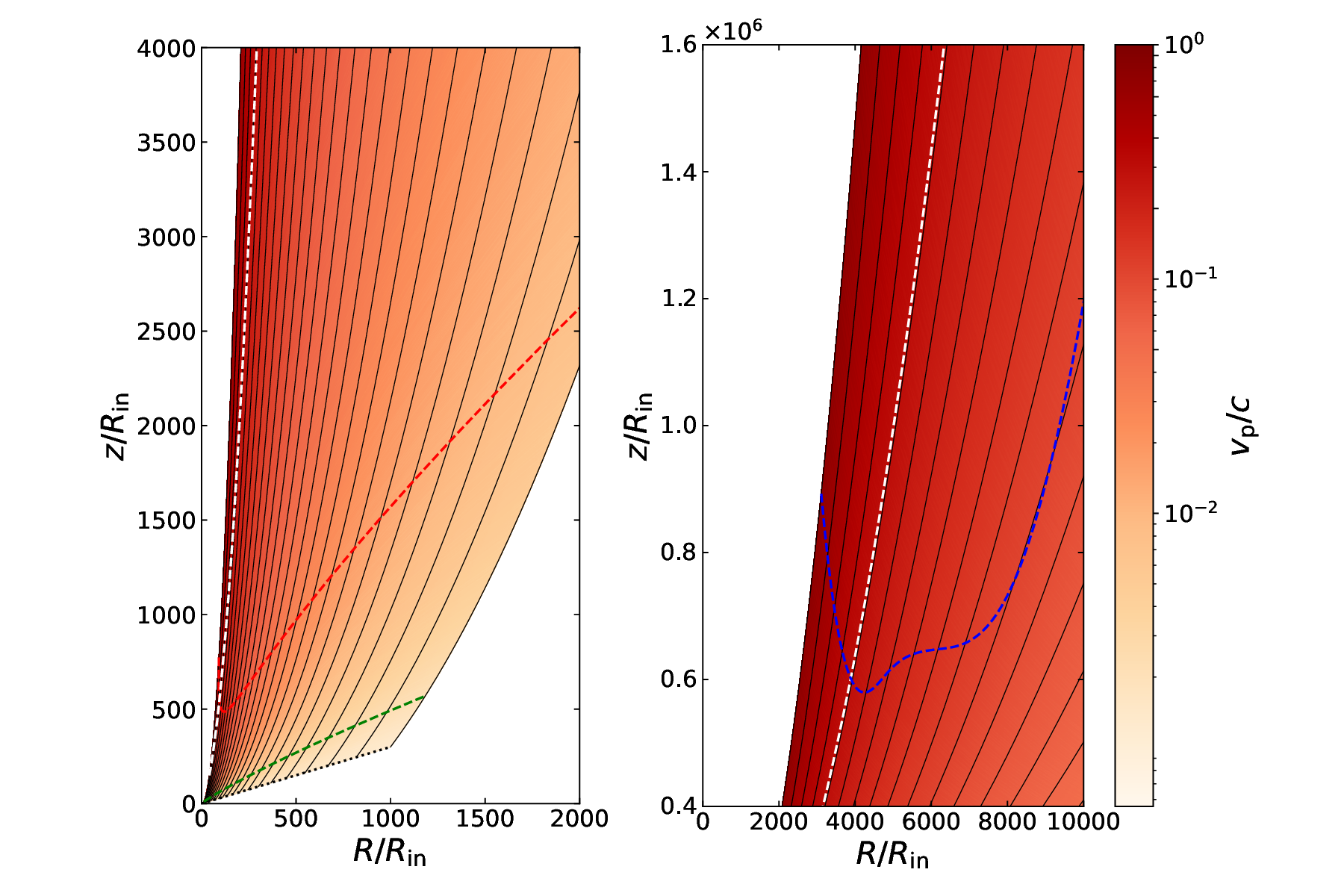}
     \includegraphics[width=0.7\textwidth]{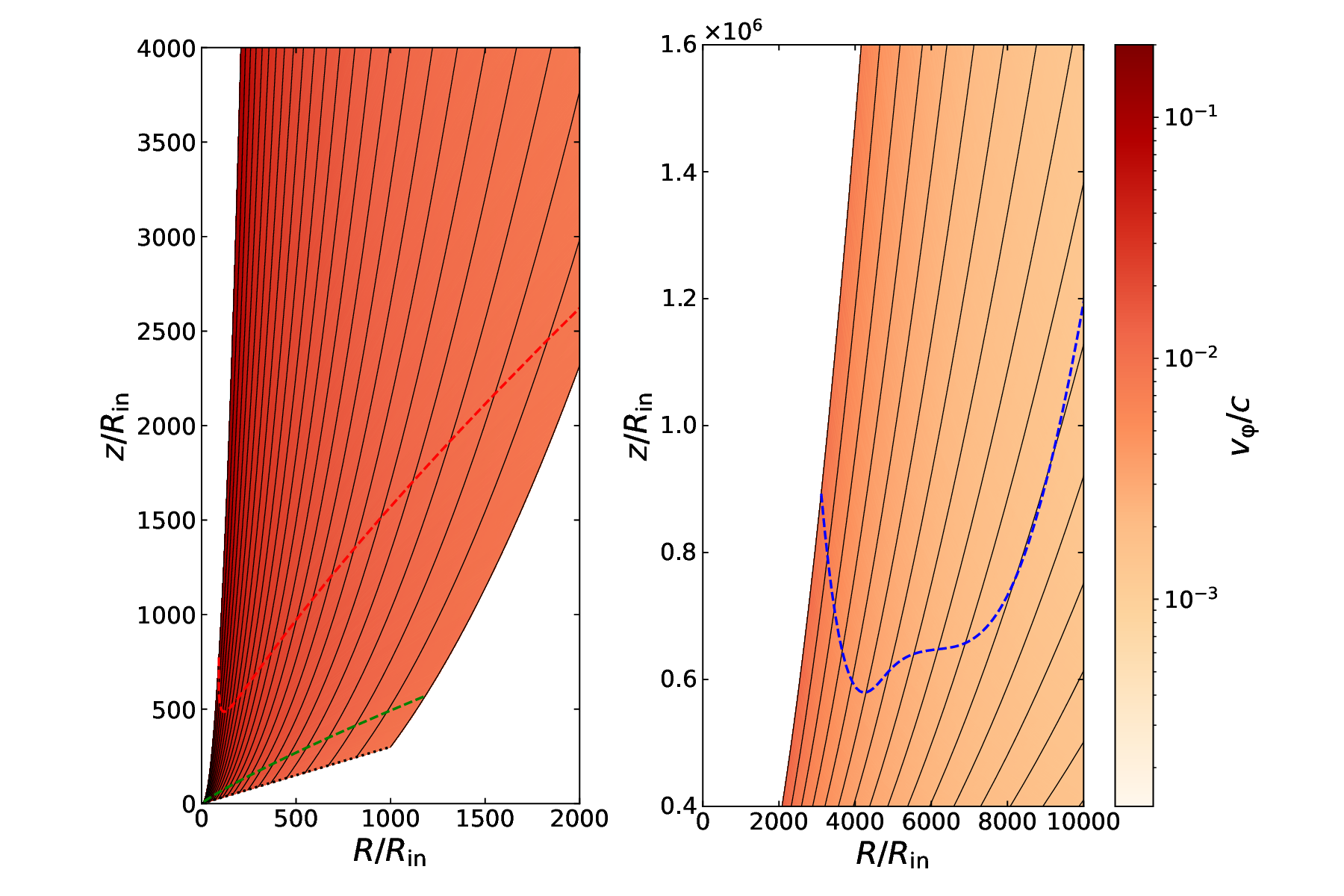} 

 \caption{The same as Figure \ref{fig:vp_vphi_tripcolor_gamma_1.4}, but calculated for $\gamma=1.5$}
    \label{fig:vp_vphi_tripcolor_gamma_1.5} 
\end{figure*}

\begin{figure*}
     \includegraphics[width=0.35\textwidth]{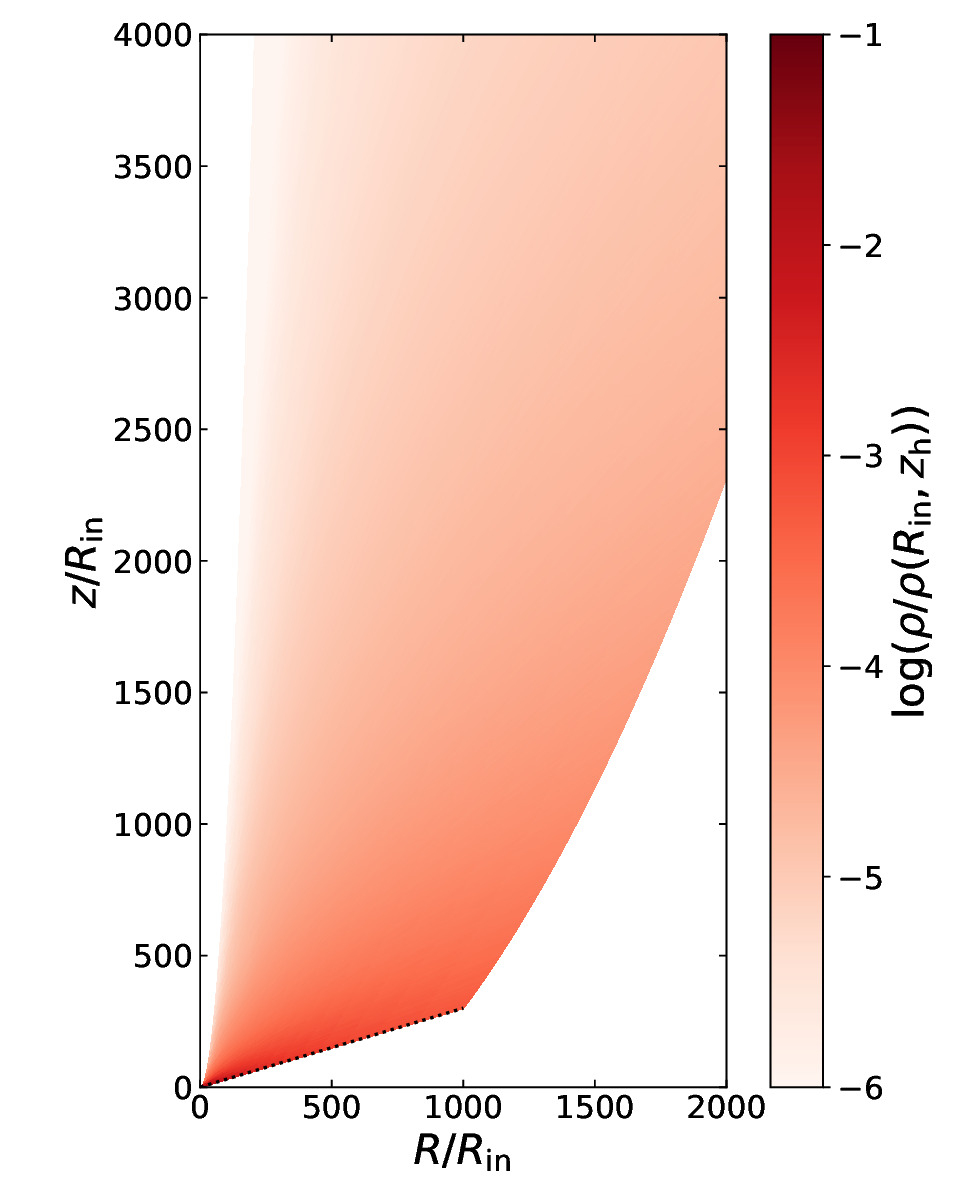}
     \includegraphics[width=0.35\textwidth]{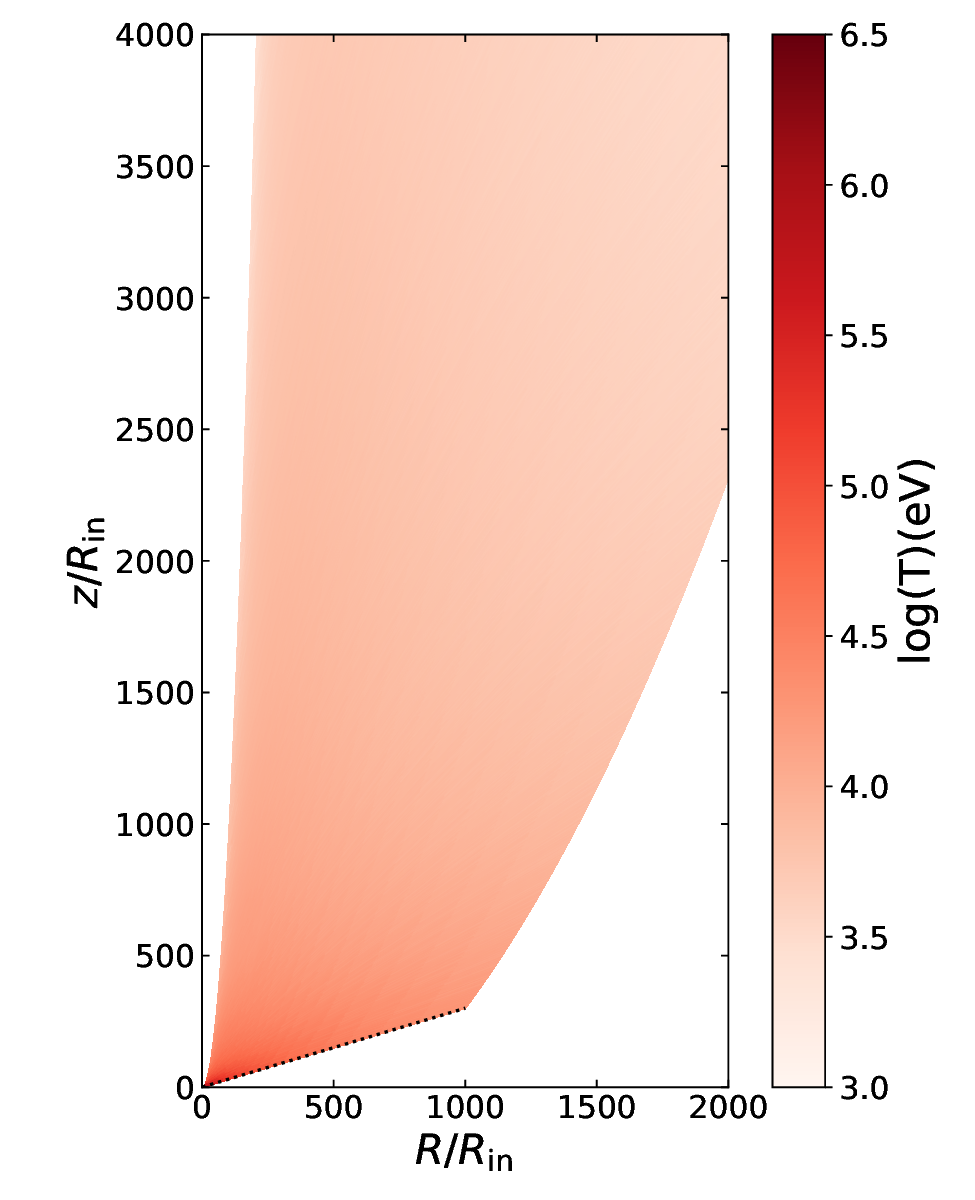} 

     \caption{The same as Figure \ref{fig:Density and temperature profiles_gamma_1.4}, but calculated for $\gamma=1.5$}  
    \label{fig:Density and temperature profiles_gamma_1.5}
\end{figure*}

\begin{figure*}
     \includegraphics[width=0.45\textwidth]{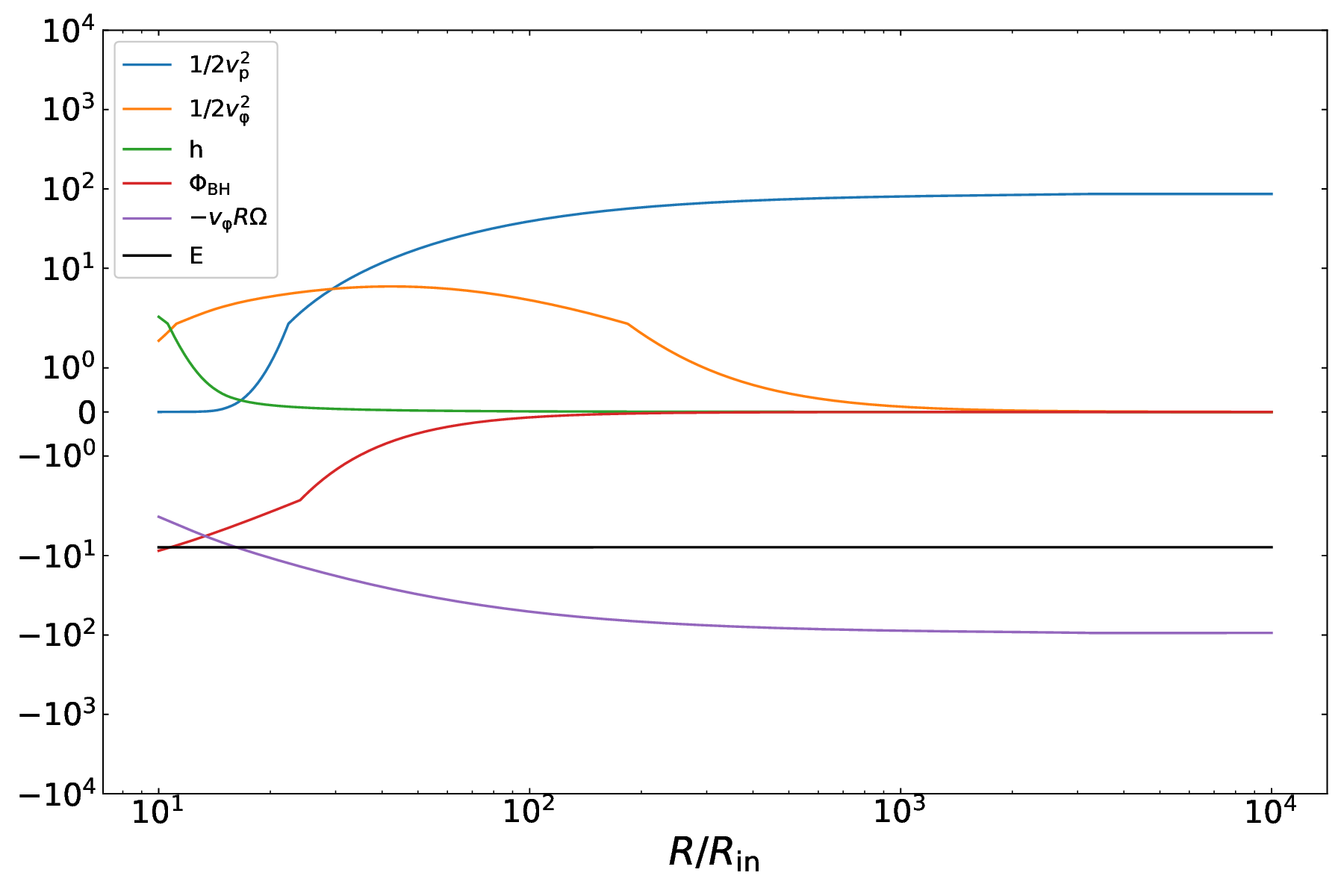}
     \includegraphics[width=0.45\textwidth]{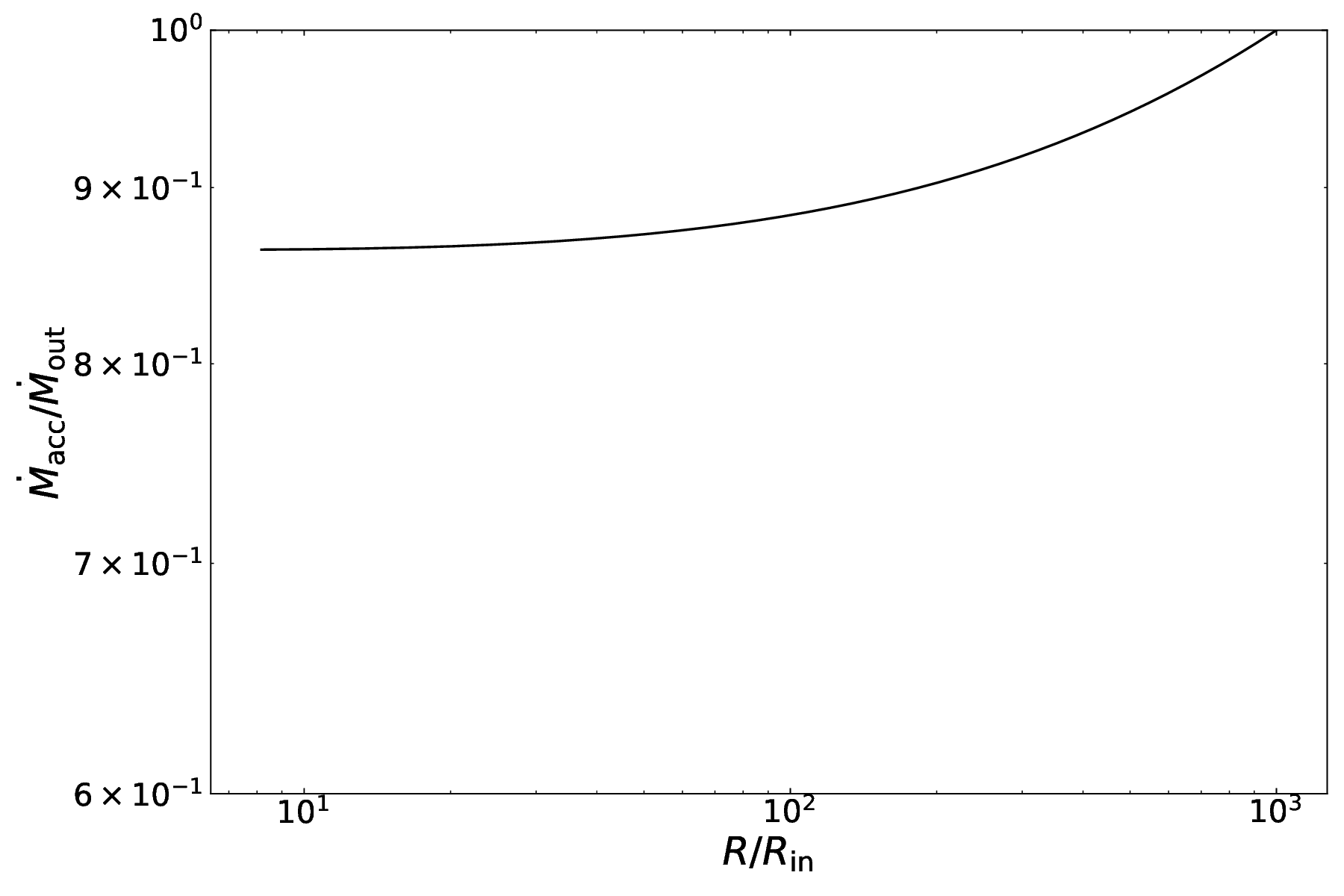} 

      \caption{the same as Figure \ref{fig:Bernoulli_conservation and mass accretion rate_gamma_1.4}, but calculated for $\gamma=1.5$}
    \label{fig:Bernoulli_conservation and mass accretion rate_gamma_1.5}
\end{figure*}

\begin{figure*}
     \includegraphics[width=0.6\textwidth]{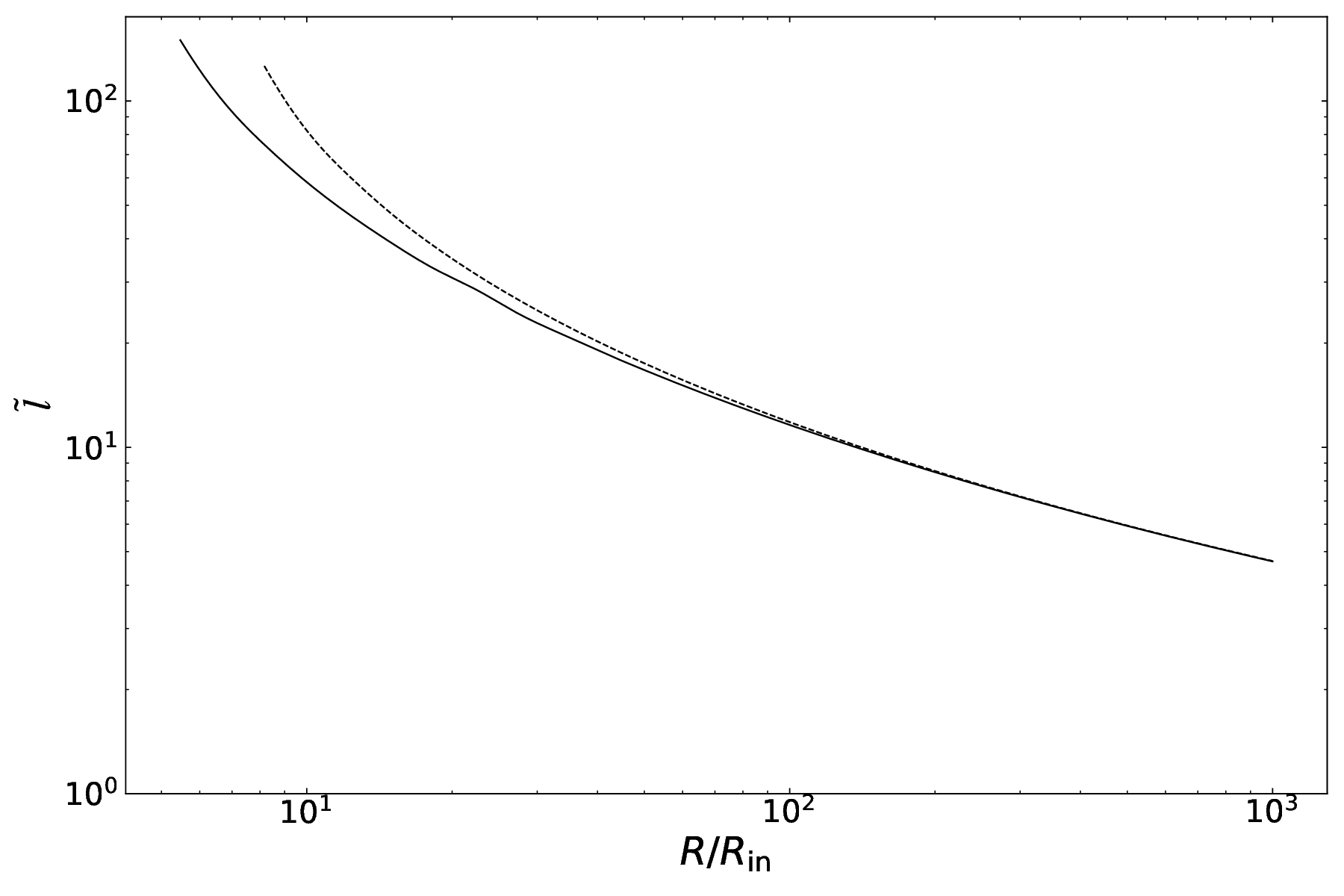}
      \caption{The dimensionless parameter $\tilde{l}=R^2_{\rm A}/R^2$ varies with radius $R$, the solid line is calculated for  $\gamma=1.4$ while dashed line is for  $\gamma=1.5$,}
      \label{fig:tilde_l}
\end{figure*}

\section{Conclusions and Discussions} \label{sec:CandD}
In this work, based on the dynamical structure of the MAD given by \cite{2022ApJ...941...30C}, we have constructed a semi-analytical model for magnetically driven outflows launched from the surface of a magnetically arrested disc. The strong large-scale magnetic field accumulated in the MAD efficiently accelerates the gas along the field lines. For the outflow driven along the inner field lines, the formal Newtonian solutions can enter the mildly relativistic regime, although the precise values of the highest velocities should not be interpreted quantitatively because relativistic corrections are not included in the present model. Our work suggests that MADs may be unified engines of both relativistic jets and UFO-like disc outflows/winds, and the present semi-analytic model provides physical insight into how the outflow component is accelerated. The strong, ordered magnetic fields expected in MADs can accelerate the outflow to velocities comparable to those inferred for fast and ultra-fast outflows in black hole accretion systems. As the disc is threaded by a dynamically important large-scale poloidal magnetic field, the trans-magnetosonic outflow solutions can be obtained. The gas in the outflow is initially launched from the disc surface, passes smoothly through the slow magnetosonic point, the Alfv\'en point, and the fast magnetosonic point, and is subsequently accelerated to sub-relativistic or mildly relativistic velocities. Near the disc surface, the gas is constrained to move along relatively stiff poloidal field lines because the magnetic pressure totally dominates the gas pressure, thus the Alfv\'en point is relatively far away from the disc surface (see the red-dashed line in Figure \ref{fig:vp_vphi_tripcolor_gamma_1.4} and \ref{fig:vp_vphi_tripcolor_gamma_1.5}), but for the weak magnetic field case, the Alfv\'en point and slow magnetosonic point are very close to each other and are both very close to the surface of the disc (see Figure 3 in \cite{2022ApJ...926...11L}), hence a relatively low terminal speed of the outflows. In a SANE disc, the large-scale magnetic flux is relatively weak (i.e., the gas pressure dominates the magnetic pressure at the main disc region), and the launching of a coherent outflow depends sensitively on the instantaneous field geometry. In a MAD, by contrast, the accumulated magnetic flux reaches a saturation level and becomes dynamically important in the inner accretion flow \citep[][]{2003PASJ...55L..69N,2011MNRAS.418L..79T,2012MNRAS.423.3083M},  the detection of powerful, highly ionized, high-velocity disc outflows may provide an indirect diagnostic of strongly magnetized accretion states. Although much of the MAD researches have focused on the production of highly relativistic Blandford-Znajek jets, recent GRMHD simulations have shown that MADs also drive substantial wide-angle outflows. \cite{2022ApJ...941...30C} found that magnetic flux eruptions in the MADs around a nonspinning black hole can push gas away from the disc and launch strong outflow, which provides an efficient vertical channel for angular momentum transport. While for the SANE model, the magneto-rotational instability (MRI) moves angular momentum mostly equatorially through the disc. More recent simulations further suggest that disc outflows can contribute an angular momentum flux transport comparable to that associated with internal turbulent stresses in thick MADs \citep[][]{2024ApJ...965..175M}. The angular momentum transport will significantly affect the dynamical structure of the accretion disc including the MAD disc, which is a key to understanding the evolution of systems where geometrically thick MADs can occur, such as the hard state of XRBs, low-luminosity AGNs, and some tidal disruption events. Our present work can just complement the numerical studies by isolating the steady magnetocentrifugal component of the MAD-driven outflow and by clarifying how the outflow velocity, density, and critical-point structure depend on the prescribed disc dynamical conditions. To what extent the fast outflow can affect the dynamical structure of the MAD disc is an important issue, which is out of the scope of  present work, and we will make a further study in the near future.

The velocity structure derived in the present work may be relevant to the fast and ultra-fast outflows observed in AGNs and X-ray binaries. Observationally inferred UFO velocities are commonly of order $0.1c$ to several tenths of the speed of light, with still higher velocities reported in some systems. In our calculations, magnetocentrifugal acceleration naturally produces velocities in the $0.1c-0.3c$ range within the regime in which the Newtonian treatment remains comparatively well controlled. The formal solutions associated with the innermost field lines continue to accelerate to still higher velocities, however, these values should be interpreted qualitatively because special-relativistic effects are not included. Such velocities are difficult to obtain by pure thermal driving, and line driving (which may become inefficient when the gas is highly ionized). Magnetically driven outflows provide an attractive alternative except pure thermal driving and line driving mechanisms, especially in systems where the inner disc is threaded by a coherent large-scale field \citep[][]{2010ApJ...715..636F,2015ApJ...805...17F,2015ApJ...814...87M,2018ApJ...857...34Z,2019ApJ...872..149L}. In our present calculations, the outflows launched from the inner region of the disc can be accelerated to velocities comparable to those inferred for the observations of UFOs, while the outflows launched from larger radii produce slower components (see Figure \ref{fig:variables} and \ref{fig:vp_vphi_tripcolor_gamma_1.4}), this naturally leads to a stratified outflow, in which the inner outflow may be identified with highly ionized UFOs and the outer outflow with lower-velocity absorbers. Such a radially stratified structure is a common feature of self-similar magnetic outflow models and may help explain the broad range of ionization parameters and velocities inferred from the observational spectra. The strong large-scale magnetic field required in the present calculation is supported by growing observational evidence for dynamically important magnetic fields near accreting black holes. The high-resolution polarimetric EHT observations of M87* favor models in which ordered magnetic fields play a major dynamical role near the event horizon \citep[][]{2021ApJ...910L..13E}, while multiwavelength observations of the X-ray binary MAXI J1820+070 have been interpreted as evidence for the formation of a MAD disc during outburst \citep[][]{2023Sci...381..961Y}. If MAD discs are common in both supermassive and stellar-mass black hole systems, magnetically driven disc winds/outflows may provide a unified mechanism for producing UFO-like absorbers across a wide range of black hole masses. As proposed in \cite{2019ApJ...872..149L}, the magnetically driven outflow is an efficient way to transfer disc angular momentum outward, which will significantly increase the radial accretion velocity of the disc. Considering the fast outflow driven from the MAD disc, we infer that outflow-driven accretion may also significantly affect dynamical evolution of the MAD disc, which may provide a mechanism for explaining the rapid variability and accretion-state transitions observed in strongly magnetized black hole systems, such as some of the changing-look AGNs and black hole X-ray binaries.

Current calculation find the large magnetic lever arm associated with the MAD-driven outflow. For the fiducial solutions, we find $\tilde{l}=(R_{\rm A}/R)^2$ to be typically several tens and, for some inner field lines, as large as $\sim10^2$. This implies that the  Alfv\'en surface lies several to more than ten launching radii away from the disc. Since the conserved specific angular momentum carried by the outflow is $l_{\rm w}=\tilde{l} l_{\rm d}$, even a relatively small amount of mass loss can remove a disproportionately large amount of angular momentum from the accretion disc. Large lever arms are a well-known possibility in magnetocentrifugal outflow theory, particularly for strongly magnetized and lightly mass-loaded solutions. For comparison, works of weakly magnetized disc outflow often found lever arms of order unity to a few, whereas strongly magnetized, weakly mass-loaded self-similar magnetocentrifugal solutions can reach $\sim 10^2$ \citep[][]{1994A&A...287...80C,2011ApJ...733...57T,2014ApJ...783...51C,2018ApJ...857...34Z,2019MNRAS.490.3112J,2021A&A...650A..35L,2022ApJ...926...11L}. Our results show that similarly large lever arms can arise naturally in a black hole MAD environment when the outflow-base conditions are constrained by the strongly magnetized GRMHD MAD structure. The MAD state provides a natural physical route to the high-magnetization, low-mass-loading regime result in such efficient magnetic angular-momentum extraction. This result also provides a more direct connection between the outflow and the dynamical evolution of the MAD. A large $\tilde{l}$ implies that the outflow torque can remain dynamically important even when the outflow mass-loss rate is only a moderate fraction of the accretion rate. The MAD outflow may therefore contribute substantially to driving accretion by vertically removing angular momentum from the disc, complementing the radial angular-momentum transport produced by Maxwell and Reynolds stresses within the accretion flow.

The comparison between the $\gamma=1.4$ and $\gamma=1.5$ cases indicates that the qualitative properties of the outflow are robust against moderate changes in the assumed polytropic index. The locations of the slow sonic point and the Alfv\'en point remain nearly unchanged. A larger $\gamma$ changes the thermal enthalpy available near the slow point and slightly modifies the density and temperature distributions, but it does not affect the basic acceleration process. The thermodynamics of the outflow has been represented by a polytropic equation of state with a constant adiabatic index, this assumption is useful for obtaining global trans-critical solutions. But we note here that in astrophysical outflowing systems, thermodynamics along the field line should be important for observational applications. The ionization state, column density, and radiative heating/cooling history of the gas will determine whether the outflow can be detected as an X-ray absorber, these properties may be studied in detail by numerical simulations. 

The present model contains several other simplifications that should be addressed in future work. The magnetic field geometry is prescribed rather than solved self-consistently with the disc structure. In a real MAD, the poloidal field geometry, toroidal field generation, disc thickness, and mass loading are tightly coupled through the nonlinear GRMHD process. The assumed field configuration provides a useful controlled approximation, but it cannot capture magnetic flux eruptions,  interchange instabilities, or non-axisymmetric accretion streams. These effects are known to be important in saturated MADs and may introduce time-dependent variability in the outflow velocity and mass flux. The present calculation is performed within a Newtonian MHD framework and therefore does not explicitly include general-relativistic effects. We emphasize, however, that the inner boundary of our outflow calculation is located at ($R_{\rm zh,min}=5R_{\rm in}=30R_{\rm g}$), and all of the outflow solutions considered in this work are launched from ($R\geq30R_{\rm g}$), the outflow-launching region lies well outside the strongly relativistic regime close to the event horizon. We therefore expect general-relativistic corrections to the gravitational potential and the outflow-launching dynamics to be relatively modest over the radial range considered in this work. This radial restriction, however, does not remove the separate issue associated with the high velocities reached by some of the calculated outflow solutions. When the formally derived velocity approaches a substantial fraction of the speed of light, special-relativistic corrections to the flow dynamics become non-negligible. The highest velocities obtained from our Newtonian solutions should therefore not be interpreted as precise quantitative predictions of the terminal velocity of the outflow. Instead, these solutions demonstrate that the strong large-scale magnetic field in the MAD regime is capable of accelerating the outflow into the mildly relativistic regime. A fully relativistic MHD treatment would be required to determine accurately the outflow terminal velocity. Our present model does not include black-hole spin, or the dynamical coupling between the disc outflow and the Blandford Znajek jet. These effects become particularly important closer to the black hole and require a fully general-relativistic treatment. GRMHD simulations have shown that rapidly spinning black holes in the MAD state can produce powerful relativistic jets through the extraction of rotational energy \citep[][]{2011MNRAS.418L..79T}. In such systems, the disc outflow and the relativistic jet should be treated as two coupled components of the same magnetic outflow system. Investigating the coupling between such a relativistic jet and the slower, wide-angle disc outflow considered here is beyond the scope of the present work. A fully relativistic treatment in the future work including Kerr geometry and relativistic MHD effects is required to determine the exact velocity and Lorentz-factor distribution of the innermost outflows.

\section*{Acknowledgements}
We thank Professor Xinwu Cao for helpful discussions and the referee for valuable comments and suggestions. This work is supported by the NSFC (12303020, 12233007), the Yunnan Fundamental Research Projects (NO.202401CF070169), the Xingdian Talent Support Plan - Youth Project, and the start-up grant from Yunnan University.

%%%%%%%%%%%%%%%%%%%%%%%%%%%%%%%%%%%%%%%%%%%%%%%%
\section*{Data Availability}

The data underlying this article will be shared on reasonable request to the corresponding author.



\bibliographystyle{mnras}
\bibliography{ms} % if your bibtex file is called example.bib

%\begin{thebibliography}{99}
%\end{thebibliography}

%%%%%%%%%%%%%%%%%%%%%%%%%%%%%%%%%%%%%%%%%%%%%%%%%%

%%%%%%%%%%%%%%%%% APPENDICES %%%%%%%%%%%%%%%%%%%%%

% \appendix

% \section{Some extra material}

% If you want to present additional material which would interrupt the flow of the main paper,
% it can be placed in an Appendix which appears after the list of references.

%%%%%%%%%%%%%%%%%%%%%%%%%%%%%%%%%%%%%%%%%%%%%%%%%%

% Don't change these lines
\bsp	% typesetting comment
\label{lastpage}
\end{document}